%% file: Expt_loss_tolerant_mdiqkd.tex
\documentclass[aps,prl,reprint,amsmath,amssymb,showpacs,superscriptaddress]{revtex4-1}

\usepackage{graphicx}
\usepackage{float}
\usepackage{longtable}
\usepackage{subfigure}
\usepackage{lineno}
\usepackage{bbm}
\usepackage{multirow}
\graphicspath{{./Figures/}}
\usepackage[pdftex,  
            pdfstartview=FitH,
            CJKbookmarks=true,
            bookmarksnumbered=true,
            bookmarksopen=true,
            colorlinks, 
            pdfborder=001,   
            linkcolor=black,
            anchorcolor=blue,
            citecolor=blue,
            urlcolor=blue
            ]{hyperref}  

\begin{document}

\title{Experimental Measurement-Device-Independent Quantum Key Distribution with Imperfect Sources}

\author{Zhiyuan Tang}
\email{ztang@physics.utoronto.ca}
\affiliation{
 Centre for Quantum Information and Quantum Control\\
 Department of Physics \& Department of Electrical and Computer Engineering\\ 
 University of  Toronto,
 Toronto, Ontario, M5S 3G4, Canada
 }

\author{Kejin Wei}
\affiliation{
 Centre for Quantum Information and Quantum Control\\
 Department of Physics \& Department of Electrical and Computer Engineering\\ 
 University of  Toronto,
 Toronto, Ontario, M5S 3G4, Canada
 }
 \affiliation{
 School of Science and State Key Laboratory of Information Photonics and Optical Communications,\\
 Beijing University of Posts and Telecommunications,
 Beijing, 100876, People's Republic of China
 }
 
 \author{Olinka Bedroya}
 \affiliation{
  Centre for Quantum Information and Quantum Control\\
 Department of Physics \& Department of Electrical and Computer Engineering\\ 
 University of  Toronto,
 Toronto, Ontario, M5S 3G4, Canada
 }
 
 \author{Li Qian}
 \affiliation{
  Centre for Quantum Information and Quantum Control\\
 Department of Physics \& Department of Electrical and Computer Engineering\\ 
 University of  Toronto,
 Toronto, Ontario, M5S 3G4, Canada
 }
 
 \author{Hoi-Kwong Lo}
 \affiliation{
  Centre for Quantum Information and Quantum Control\\
 Department of Physics \& Department of Electrical and Computer Engineering\\ 
 University of  Toronto,
 Toronto, Ontario, M5S 3G4, Canada
 }


\date{\today}

\begin{abstract}
Measurement-device-independent quantum key distribution (MDI-QKD), which is immune to all detector side-channel attacks, is the most promising solution to the security issues in practical quantum key distribution systems. Though several experimental demonstrations of MDI-QKD have been reported, they all make one crucial but not yet verified assumption, that is there are no flaws in state preparation. Such an assumption is unrealistic and security loopholes remain in the source. Here we present, to our knowledge, the first MDI-QKD experiment with the modulation error taken into consideration. By applying a security proof by Tamaki \textit{et al} (Phys. Rev. A 90, 052314 (2014)), we distribute secure keys over fiber links up to 40 km with imperfect sources, which would not have been possible under previous security proofs.  By simultaneously closing loopholes the detectors and a critical loophole - modulation error in the source, our work shows the feasibility of secure QKD with practical imperfect devices.
\end{abstract}

\pacs{03.67.Dd, 03.67.Hk, 42.50.Ex}

\maketitle


Quantum key distribution (QKD), in principle, offers unconditional security based on the laws of quantum physics rather than computational complexity \cite{BB84,*PhysRevLett.67.661}. However, it has been realized that, due to the gap between the security proof and real-life implementations, practical QKD systems are vulnerable to various attacks \cite{PhysRevA.78.042333, *NaturePhotonics, *NatureComm, *PhysRevLett.107.110501, *NJP.13.073024, *NJP.12.113026}. 

Device-independent QKD (DI-QKD) \cite{DIQKD, *PhysRevLett.98.230501, *PhysRevLett.105.070501, *PhysRevLett.113.140501}, was proposed to remove all assumptions of the internal working of devices of QKD. The security of DI-QKD is based on the loophole-free Bell test. Despite a number of recent experimental demonstrations of loophole-free Bell test \cite{hensen2015loophole, *PhysRevLett.115.250401, *PhysRevLett.115.250402}, DI-QKD is impractical at practical distances (20-30 km of telecom fiber) due to its low key rate of about $10^{-10}$ bit per pulse \cite{PhysRevA.84.010304}. Fortunately a protocol, namely the Measurement-Device-Independent QKD (MDI-QKD), whose security is built on the time-reversed entanglement QKD \cite{PhysRevA.54.2651, *Algorithmica.34.340} , has been proposed \cite{PhysRevLett.108.130503} to remove all potential security loopholes in the detection side, the most vulnerable part of a QKD system (See also \cite{PhysRevLett.108.130502}).  Several MDI-QKD demonstrations using polarization \cite{PhysRevLett.112.190503, PhysRevA.88.052303}  and time-bin phase \cite{PhysRevLett.111.130501, *PhysRevLett.111.130502} encoding have been reported. More recently, MDI-QKD over 200 km \cite{PhysRevLett.113.190501}, a field test \cite{IEEE.J.Sel.T.Quantum.Electronics.21.6600407}, a network demonstration \cite{2015arXiv150908389T}, and an implementation with 1 GHz clock rate \cite{2015arXiv150908137C} have been reported, highlighting the practicality of this protocol. MDI-QKD with continuous variables has also been proposed in \cite{pirandola2015high, *PhysRevA.89.052301, *PhysRevA.89.042335}.

It is conceivable that MDI-QKD \cite{PhysRevLett.108.130503} will be widely adopted in the near future. Since MDI-QKD is intrinsically immune to all detector side-channel attacks, eavesdroppers will shift their focus from hacking the detectors to hacking the sources, which are not protected in MDI-QKD. Several theoretical studies on MDI-QKD with imperfect sources have been reported \cite{PhysRevA.88.062322,*PhysRevA.90.052319,*PhysRevA.92.012333}.

A crucial assumption in discrete-variable MDI-QKD  is that the source employed must be trusted. An ideal trusted source need to satisfy two conditions: first, the source only emits single photons; second, information should be encoded without flaws. However, these two conditions cannot be satisfied perfectly with today's technology. First, phase-randomized weak coherent pulses (WCPs) rather than single-photon sources are widely used in most QKD (including BB84 and MDI-QKD) demonstrations. Fortunately, it has been shown that unconditional security can still be achieved with phase-randomized WCPs \cite{GLLP}. Furthermore, the performance can be significantly improved with the decoy state method \cite{PhysRevLett.91.057901, *PhysRevLett.94.230504, *PhysRevLett.94.230503, *PhysRevLett.96.070502, *PhysRevLett.98.010503, *PhysRevLett.98.010505}. Second, encoding quantum states onto optical pulses has inherent errors due to the finite inaccuracies in practical encoding devices. However, such errors are ignored in all previous discrete-variable MDI-QKD demonstrations \cite{PhysRevLett.112.190503,PhysRevLett.111.130501, PhysRevLett.111.130502, PhysRevA.88.052303, PhysRevLett.113.190501, IEEE.J.Sel.T.Quantum.Electronics.21.6600407}. It is unrealistic to ignore all those errors because they may lead to security loopholes that a eavesdropper might conceivably exploit to launch attacks.

Such state preparation flaws can be taken care of using the quantum coin idea \cite{GLLP, PhysRevA.85.042307}. However, this approach assumes the worst case in which an eavesdropper can enhance the flaws by channel loss, and therefore the performance is not loss tolerant. The study in \cite{PhysRevA.85.042307} shows that highly accurate state preparations are required in MDI-QKD.

Recently,  Tamaki \textit{et al} have proposed a loss-tolerant security proof \cite{PhysRevA.90.052314} that can take modulation error - a most crucial flaw in a QKD source, into consideration. The loss-tolerant protocol is secure against the most general type of attacks. For ease of discussion, the intuition behind the security of the loss-tolerant protocol can be understood for the example of the unambiguous state discrimination (USD) attack. The idea is that, as long as the states are encoded in 2-dimensional qubits \footnote{Similarly, the bounded dimensionality of the encoding space is also assumed in the semi-device independent QKD \cite{PhysRevA.84.010302}. However loss-tolerant MDI-QKD gives a much higher key rate than semi-device independent QKD and is thus more practical. }, it is impossible for Eve to launch a USD attack. Therefore Eve cannot enhance state preparation flaws of qubits by channel loss. The performance of QKD can thus be dramatically improved even when the state preparation flaws are considered. This idea has been applied to both the BB84 protocol and the three-state prepare-and-measure protocol \cite{1504.08151},  and an experimental demonstration is reported in \cite{1408.3667}.

It is noteworthy that this security proof can be applied to MDI-QKD. In this Letter, we extend the work in \cite{PhysRevA.90.052314} and present an experimental demonstration of MDI-QKD with state preparation imperfections over fiber links of 10 km and 40 km.  By closing an important potential loophole in MDI-QKD, we achieved improved security compared to previous demonstrations.

The contributions of this Letter are as follows. First and most importantly, in contrast to previous MDI-QKD demonstrations \cite{PhysRevLett.112.190503,PhysRevLett.111.130501, PhysRevLett.111.130502, PhysRevA.88.052303, PhysRevLett.113.190501, IEEE.J.Sel.T.Quantum.Electronics.21.6600407} which unrealistically assume perfect state preparations, we carefully optimize the state preparation to minimize the preparation flaws and perform a complete characterization of the states using quantum state tomography. For the first time, we include the state preparation flaws into secure key rate estimation. We highlight that this would not have been possible under previous security proofs \cite{GLLP, PhysRevA.85.042307}. Second, we remark that the analysis in \cite{PhysRevA.90.052314} only applies to the asymptotic case with an infinite number of decoy states and an infinitely long key. We here present the theory (see Supplemental Material) which shows how the loss-tolerant protocol can be applied to MDI-QKD in a realistic setting, where only a finite number of decoy states and a key of finite length are available. Third, we improve the key generation speed by increasing the system repetition rate from 500 kHz \cite{PhysRevLett.112.190503} to 10 MHz and employing free-running single-photon detectors with 20\% quantum efficiency.  These technological improvements enable us to get a positive key rate within a reasonable time frame, even when finite key effects and encoding flaws are taken into account, and thus demonstrate the practicality of the protocol. 

We first briefly explain the loss-tolerant MDI-QKD protocol.  Alice (Bob) randomly encodes her (his) key bits into one of the three polarization states \{$\rho_{0_Z}$, $\rho_{1_Z}$, $\rho_{0_X}$\}, where $\rho_{i_\alpha}$ is the density matrix of the polarization state of single photons corresponding to the bit value $i\in\{0, 1\}$ in the basis $\alpha\in \{Z, X\}$. She (He) then sends her (his) encoded WCPs to an untrusted third party, Eve, who can be an eavesdropper, to do Bell state measurements (BSMs). After a sufficient number of key bits have been transmitted, Eve announces the BSM results to Alice and Bob. Alice and Bob also announce their basis choices over a public authenticated channel and generate a sifted key. By revealing part of the sifted key, they can estimate the bit error rate in the $Z$ basis and perform error corrections. 

We apply the decoy state method \cite{NJP.15.113007} to estimate the gain of single photons in the $Z$ basis. The phase error rate of single photons $e^X_{11}$, which quantifies the information leakage to an eavesdropper,  is estimated from the transmission rates of fictitious states using the rejected data analysis \cite{PhysRevA.90.052314}. Privacy amplification can then be performed to generate a secret key.

\begin{figure}[htbp]
	\centering
	\includegraphics[width=0.5\textwidth]{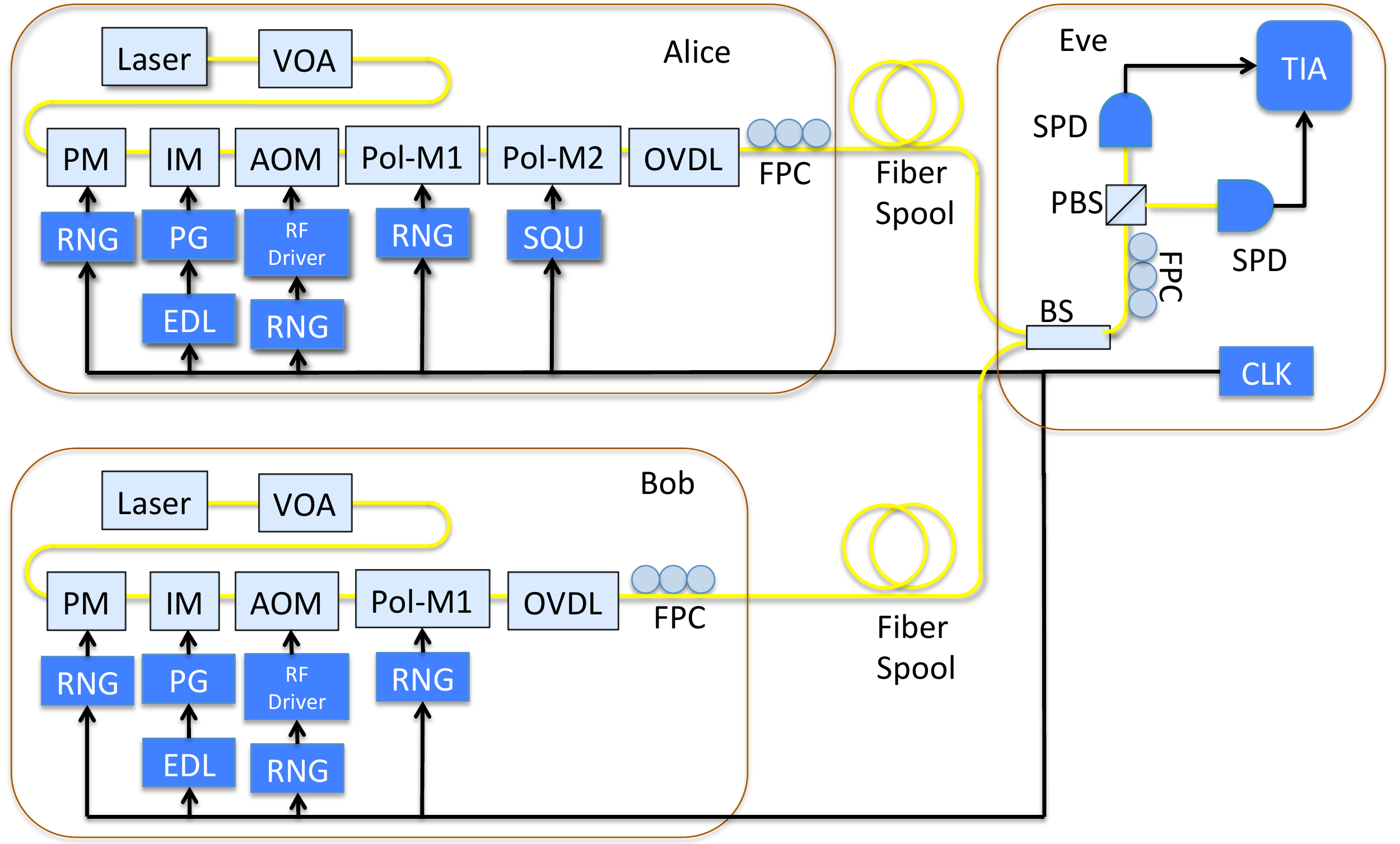}
	\caption{\label{expt_setup}(Color online). Schematic of the experiment. Alice (Bob) possesses a frequency-locked CW laser at 1542 nm. The light is attenuated by an variable optical attenuator (VOA). The phase is randomly modulated by a phase modulator (PM) and the intensity is modulated by an intensity modulator (IM) to generate phase-randomized WCPs. An acousto-optic modulator (AOM) is used to randomly switch the intensity between signal and decoy states. Key bits are encoded by a polarization modulators (Pol-M1). Pol-M2 in Alice is used for polarization alignment. Eve's Bell state measurement setup consists of  a beam splitter (BS), three fiber polarization controllers (FPCs), a polarizing beam splitter (PBS), two single-photon detectors (SPDs), and a time interval analyzer (TIA). Acronyms for other components: RNG: random number generators; PG: pulse generators; EDL: electrical delay line; SQU: square wave generator; OVDL:  optical variable delay line.}
\end{figure}

Fig.\ref{expt_setup} shows the schematic of our experiment. Alice and Bob each have a CW laser whose wavelength is independently locked to the P16 line of a C13 acetylene gas cell (integrated in Alice's and Bob's lasers by the manufacturer) at 1542.38 nm. The frequency locking ensures that the frequency difference between Alice's and Bob's lasers is within 10 MHz, guaranteeing the spectral indistinguishability. The laser light is attenuated by a variable optical attenuator (VOA) down to single-photon level at the output of Alice's / Bob's system. Its phase is randomized by a phase modulator into 1000 discrete random phases distributed uniformly in $[0, 2\pi]$, which gives performance close to the case of continuous phase randomization \cite{NJP.17.053014}. The amplitude of the light is modulated by an intensity modulator (IM) to generate phase-randomized weak coherent pulses  at a repetition rate of $f = 10$ MHz, with a full width at half maximum (FWHM) of around 2.5 ns.

Each pulse's intensity is randomly modulated by an acousto-optic modulator (AOM). We implement the 2-decoy protocol, i.e., each pulse's amplitude is modulated to either the signal state or one of the two decoy states.

Key bits are encoded into the polarization states of the optical pulses by a polarization modulator (Pol-M). The Pol-M consists of a phase modulator, an optical circulator, and a Faraday mirror. Polarization modulation is achieved by bi-directional modulation of  the phase difference of the TE and TM components of the waveguide in the phase modulator. Details of the Pol-M setup can be found in \cite{PhysRevLett.112.190503, NJP.11.095001}. In the three state protocol, each pulse's polarization is randomly modulated to one of the three BB84 states: the horizontal state $\rho_{0_Z}$,  the vertical state $\rho_{1_Z}$, and the diagonal state $\rho_{0_X}$. We fine tuned the voltages on the Pol-Ms to minimize the preparation flaws of these states. See the Supplemental Material for details.

Alice's and Bob's pulses are sent through 2 separate fiber spools to Eve for Bell state measurements (BSMs). BSMs require indistinguishability between Alice and Bob's pulses in all degrees of freedom (except polarization, which is used for encoding). The spectral indistinguishability can be guaranteed by frequency locking in the laser as discussed above (the frequency difference of 10 MHz is much less than the bandwidth of a transform-limited pulse of 2.5 ns). To achieve the temporal indistinguishability, arrival times of Alice's and Bob's pulses are controlled by two passive electrical delay lines (EDLs) and two optical variable delay lines (OVDLs). The EDLs, which can adjust the delay of the the clock signal driving the intensity modulators (and thus the arrival time of the pulses), have a resolution of 0.5 ns and a range of 63.5 ns, and are used for coarse temporal alignments. The relative delay is further finely adjusted by the OVDLs with a resolution less than 10 ps, which is much smaller than the pulses' width of 2.5 ns FWHM.  
 
Alice and Bob need to establish a common polarization reference frame. To achieve this, they first align their $Z$ basis ($\rho_{0_Z}$ and $\rho_{1_Z}$) to the polarizing axes of the PBS in Eve's BSM setup. Alice has an extra polarization modulator (Pol-M2 in Fig.\ref{expt_setup}) in her lab. This modulator modulates the relative phase between $|H\rangle$ ($\rho_{0_Z}$) and $|V\rangle$ ($\rho_{1_Z}$). This is equivalent to a unitary rotation about the $H-V$ axis on the Poincar\'e sphere, and the amount of rotation depends on the voltage applied on Pol-M2. Alice adjusts the voltage such that her diagonal state $\rho_{0_X}$ is aligned to that of Bob.

 Alice and Bob's pulses interfere at the 50/50 beam splitter and are sent to a polarizing beam splitter (PBS), whose outputs are connected to two free running InGaAs/InP single-photon detectors (SPDs, ID220) with 20\% quantum efficiency and a dark count rate of 2 kHz. Times of the detection events (relative to the clock signal) are recorded by a time interval analyzer (TIA).  Within each period (100 ns), a 7 ns  window is chosen (by calibrating the arrival times of optical pulses) to post-select detection events.  Therefore, over 90\% of the dark count noise can be removed and the effective dark count probability per window is around $1.5\times10^{-5}$. A coincidence between these two detectors implies a successful projection onto the triplet Bell state $|\Psi^+\rangle=(|HV\rangle+|VH\rangle)/\sqrt{2}$.

We characterize the polarization states $\rho_{0_Z}, \rho_{1_Z}, \rho_{0_X}\}$ prepared by the Pol-Ms using quantum state tomography. We perform projective measurements by sending the polarization-encoded photons to a polarization analyzer (HP8169A), which consists of a half-wave plate (HWP), a quarter-wave plate (QWP), and a polarizer (POL). Angles of the waveplates and the polarizer are driven by electrical motors with an accuracy of $\pm0.1^{\circ}$ (specified by the manufacturer). A SPD is connected to the output of the polarizer for detections. Each input state $\rho_{j_\alpha}$, $j_\alpha \in \{0_Z, 1_Z, 0_X\}$, is projected into the following polarization states: $|H\rangle$ (horizontal), $|V\rangle$ (vertical), $|D\rangle$ (diagonal), and $|R\rangle$ (right-hand circular), and counts are accumulated for 10 s for each projective measurement. Density matrices can then be reconstructed using the maximum likelihood technique \cite{QIC.3.503}.

\begin{figure}[htbp]
	\centering
	\includegraphics[width=0.4\textwidth]{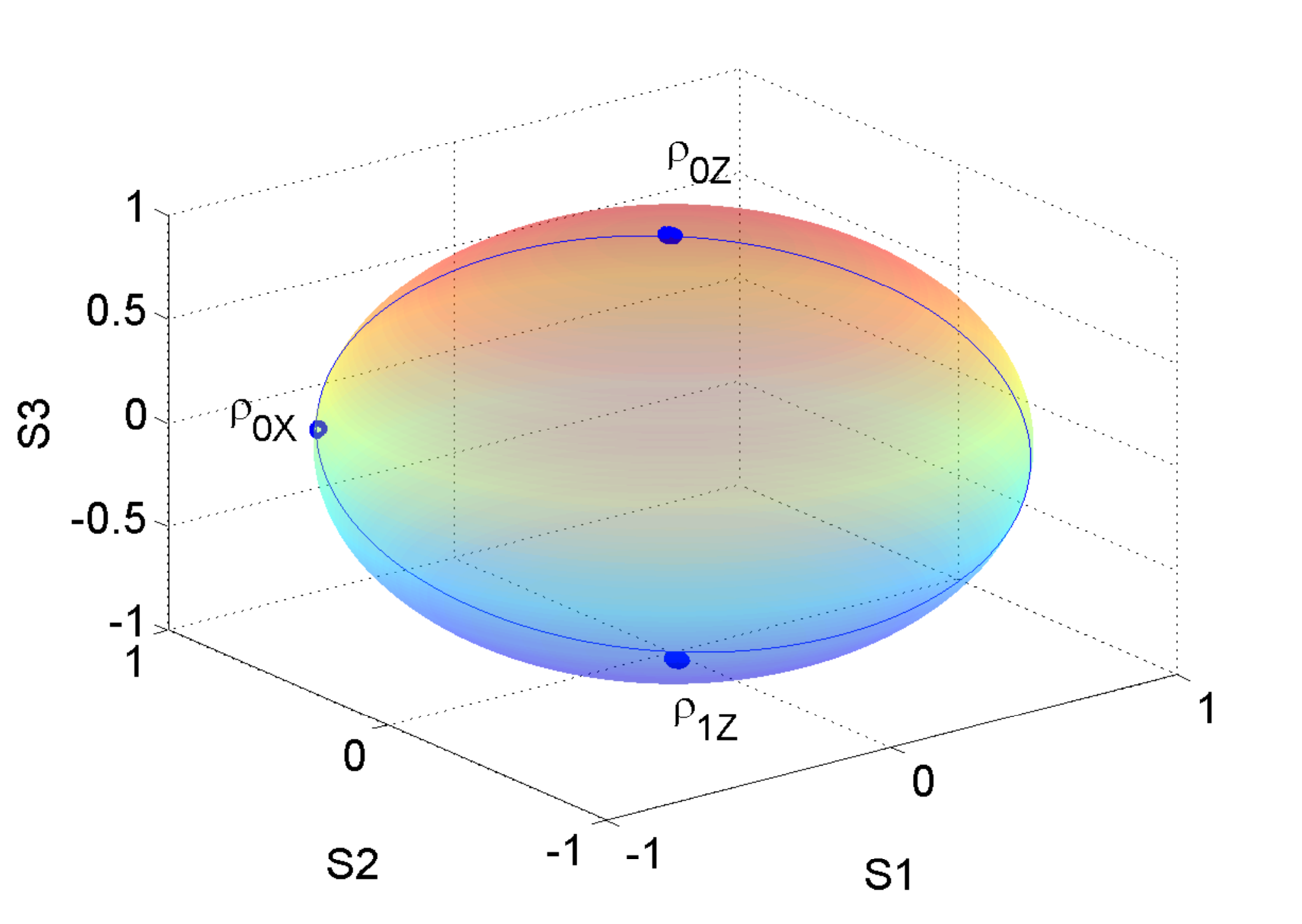}
	\caption{\label{tomography_result}(Color online).Results of quantum state tomography. Density matrices are represented by their Stokes parameters and plotted on the Poincar\'e sphere. The Stokes parameters $(S_1, S_2, S_3)$ of the states are: 
		$\rho_{0_Z}$ (-0.0032 $\pm$0.0042, 0.0106 $\pm$0.0055, 0.9994 $\pm$0.0002); 
		$\rho_{1_Z}$ (-0.0375 $\pm$0.0040, -0.0662 $\pm$0.0052, -0.9962 $\pm$0.0005);
		$\rho_{0_X}$ (-0.6963 $\pm$ 0.0028,  0.7163 $\pm$ 0.0016, -0.0128 $\pm$ 0.0029). 
	}
\end{figure}

Errors in the quantum state tomography are mostly due to the following factors: errors in counting statistics, errors in the projection states, and drift of the source's intensity and drift of the input state. We monitor the intensity during the experiment, and do not observe significant drift in intensity. The drift in input states is due to the random unitary transformation induced by the short fiber connecting the encoding system and the polarization analyzer. We characterize the stability and find that the input states remain relatively stable within the span of the quantum state tomography measurement. We therefore only consider the first two errors. Errors in counting statistics follow the Poisson distribution. Errors in projection states are due to errors in setting waveplates' angles, which follow the Gaussian distribution with an accuracy of $\pm 0.1^{\circ}$. We use Monte-Carlo method \cite{QIC.3.503} to estimate the errors in the density matrices.  Additional sets of data are generated numerically using the above distributions. Each set of data (consisting of counts and waveplate angles) is used to generate a density matrix by the maximum likelihood technique. We generate 1,000 additional simulated results for each state $\rho_{j_{\alpha}}$ to get the error distributions of the Stokes parameters. The reconstructed density matrices together with their errors are shown in Fig.\ref{tomography_result}.

We quantify the overlap between two states $\rho_{j_\alpha}$ and $\rho_{s_\beta}$ by $F(\rho_{j_\alpha}, \rho_{s_\beta})^2$, where $F(\rho_{j_\alpha}, \rho_{s_\beta}) = Tr[\sqrt{\sqrt{\rho_{j_\alpha}}\rho_{s_{\beta}}\sqrt{\rho_{j_\alpha}}}]$ is the fidelity between  $\rho_{j_\alpha}$ and $\rho_{s_\beta}$. The overlap between the states $\rho_{0_Z}$ and $\rho_{1_Z}$ is $F(\rho_{0_Z}, \rho_{1_Z})^2 = 0.0024\pm0.0006$ (whereas the ideal overlap is 0), and the overlaps between $\rho_{0_X}$ and $\rho_{0_Z}$, and between $\rho_{0_X}$ and $\rho_{1_Z}$, are $F(\rho_{0_X}, \rho_{0_Z})^2 = 0.4994\pm0.0030$ and $F(\rho_{0_Z}, \rho_{1_Z})^2 = 0.4963\pm0.0028$, respectively (whereas the ideal overlaps are 0.5). These results are comparable to other reported results in commercial \cite{1408.3667} and research \cite{JMO.62.1141} QKD systems.  Further details of the state characterization can be found in the Supplemental Material.

We implement the three state loss-tolerant MDI-QKD over 10 km and 40 km of SMF-28 optical fibers.

In the 10 km demonstration, Alice and Bob are each connected to Eve by a 5 km fiber spool. We optimize the intensities and probability distributions of the signal and decoy states using the model in \cite{NJP.15.113007}. The intensity of the signal state is chosen to be $\mu = 0.20$ photon per pulse, and the intensities for the two decoy state are $\nu_1 = 0.03$ and $\nu_2 = 0$ photon per pulse. The probability to send out the signal state $\mu$ and the decoy states $\nu_1$ and $\nu_2$ are $P_\mu=0.3$, $P_{\nu_1}=0.4$, and $P_{\nu_2}=0.3$, respectively. The probabilities to send out the states $\rho_{0_Z}$, $\rho_{1_Z}$, and $\rho_{0_X}$ are $P_{0_Z} = 0.25$, $P_{1_Z} = 0.25$, and $P_{0_X} = 0.5$, respectively. A total of $N=6\times10^{11}$ pulses are sent out. 

The lower bound of the secure key rate is given by \cite{PhysRevLett.108.130503}
\begin{equation}
\label{keyrate_eqn}
R\geq Q^{11,L}_{Z}[1-h(e^{11,U}_{X})]-Q^{\mu\mu}_Zf(E^{\mu\mu}_Z)h(E^{\mu\mu}_Z),
\end{equation}
where $Q^{11,L}_{Z}$ is the lower bound of the gain of single-photon states given that both Alice and Bob send out signal states $\mu$ in the $Z$ basis,  $e^{11,U}_{X}$ is the upper bound of the phase error rate of single-photon components, $Q^{\mu\mu}_Z$ is the gain when both of them send signal states, $E^{\mu\mu}_Z$ is the quantum bit error rate (QBER) of the signal states in the $Z$ basis, $f(E^{\mu\mu}_Z)=1.16$ is the efficiency of error correction, and $h(x)=-xlog_2(x)-(1-x)log_2(1-x)$ is the Shannon entropy. The values of  $Q^{\mu\mu}_Z$ and  $E^{\mu\mu}_Z$ are directly measured from the sifted key, and are shown in Table \ref{key_rate}.

The value of $Q^{11,L}_{Z}$ is estimated using the decoy state method \cite{PhysRevA.86.052305, NJP.15.113007}. We consider 3 standard deviations of statistical fluctuations for finite key analysis, and find  $Q^{11,L}_{Z}=3.96\times10^{-5}$. 

With the Stokes parameters of the encoded states, we upper bound the phase error rate $e_X^{11,U}=18.9\%$ using the rejected data analysis \cite{PhysRevA.90.052314} and the decoy state method.  We can then lower bound the secure key rate $R\geq2.48\times10^{-6}$ bit per signal pulse. The number of pulses where both Alice and Bob send signal states $\mu$ in the $Z$ basis is $N^{\mu\mu}_{Z}=1.35\times10^{10}$, and  a private key of length $L=N^{\mu\mu}_{Z}  R = 33.8$ kbits is generated.

\begin{table*}[!htbp]
	\caption{\label{key_rate} Key rate for loss-tolerant MDI-QKD at 10 km and 40 km. An infinitely long key ($\infty$ in data size) indicates that finite key effect is not considered when estimating the key rate $R$. The inefficiency of error correction is chosen to be $f(E_{\mu\mu}^Z)=1.16$.}
	\begin{ruledtabular}
		\begin{tabular}{cccccccc}
			Distance & Data size & Security bound & $Q_{11}^{Z, L}$ & $e_{11}^{X, U}$ & $Q_{\mu\mu}^Z$ & $E_{\mu\mu}^Z$ & $R$ (bit per pulse) \\
			\hline
			10 km & $6\times10^{11}$ & $10^{-3}$ & $3.96\times10^{-5}$ & 0.189 & $6.31\times10^{-5}$ & 0.0178 & $2.48\times10^{-6}$ \\
			
			10 km & $\infty$ & N/A & $4.17\times10^{-5}$ & 0.079 & $6.31\times10^{-5}$ & 0.0178 & $1.57\times10^{-5}$ \\
			
			40 km & $\infty$ & N/A & $1.88\times10^{-5}$ & 0.122 & $2.94\times10^{-5}$ & 0.0368 & $1.00\times10^{-6}$ \\
			
		\end{tabular}
	\end{ruledtabular}
\end{table*}

The high phase error rate is due to the small key size in this demonstration. We also estimate the key rate without finite-key correction, as shown in Table \ref{key_rate}.  Besides, we perform a proof-of-principle demonstration at 40 km. Intensities of the signal and decoy states are the same as those in the 10 km demonstration. The key rate is estimated without finite-key correction. See Table \ref{key_rate} for details.

\begin{figure}[!htbp]
 	\centering
 	\includegraphics[width=0.5\textwidth]{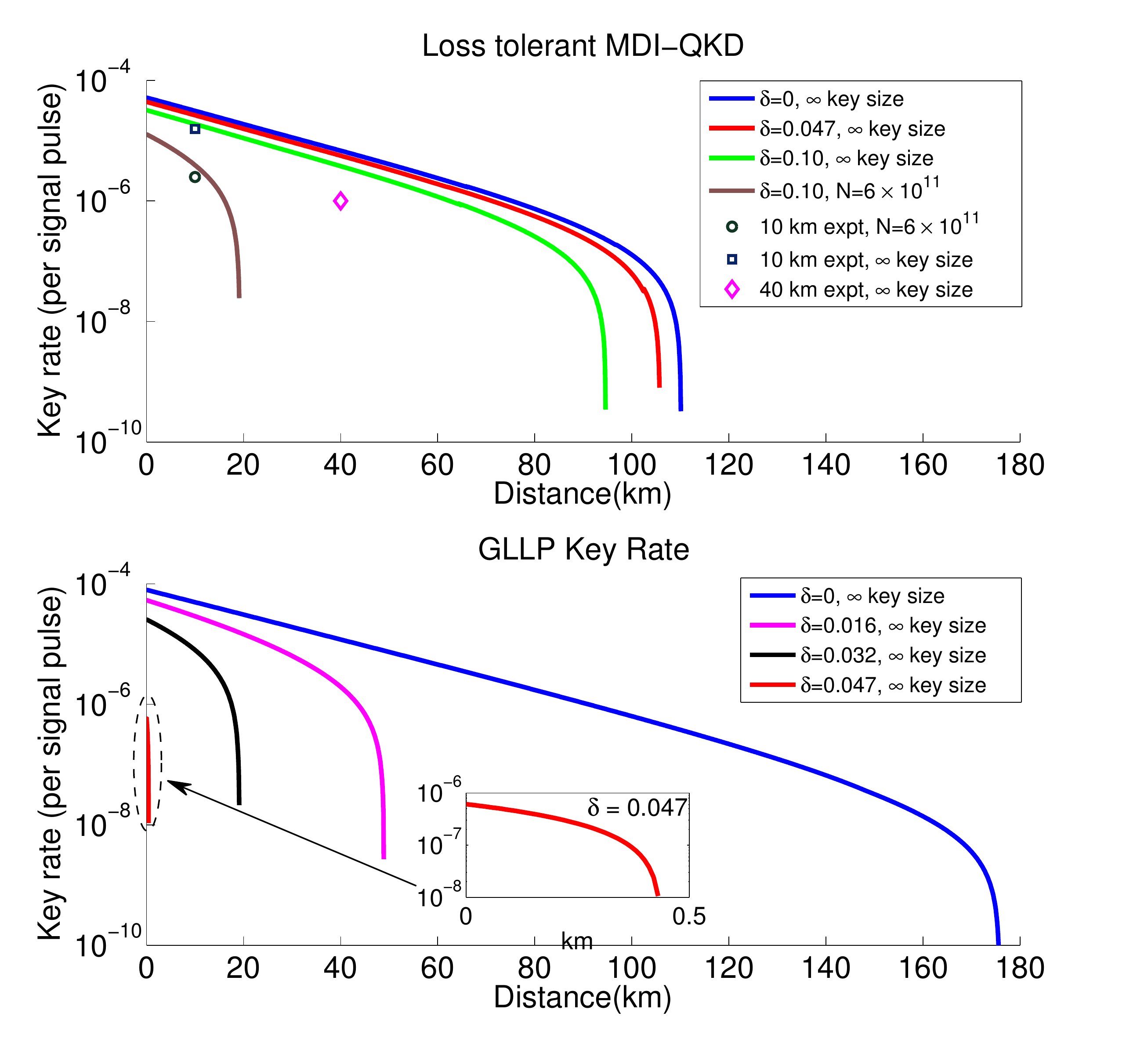}
 	\caption{\label{gllp_key}(Color online). The upper figure shows the simulated and experimental key rates of the loss-tolerant MDI-QKD protocol, for both the infinitely long key case and the finite-key case. We use $\frac{\delta}{\pi}$ to quantify the relative modulation error. See Supplemental Material for the exact definition of $\delta$. The modulation error $\delta = 0.1$ corresponds to $F(\rho_{0_Z}, \rho_{1_Z})=0.0025$, which is close to our experimental value.  The lower figure shows the simulated key rates for an infinitely long key under the GLLP analysis. The results show that the loss-tolerant protocol gives a positive key rate for realistic values of encoding flaws, while no key can be generated with the GLLP proof.  We use our experimental parameters for simulation.}
 \end{figure}

As a comparison, we simulate the performance of MDI-QKD with source flaws using the three-state loss-tolerant analysis and the GLLP analysis\cite{GLLP, PhysRevA.85.042307}. The result is shown in Fig.\ref{gllp_key}, which indicates that no secure key can be generated using the GLLP analysis, even for an infinitely long key.

In summary, we have demonstrated the first MDI-QKD experiment with an important type of source flaws taken into consideration.  In contrast to previous demonstrations which assume perfect state modulations without verification, our experiment shows the feasibility of generating secure keys with imperfect states prepared by off-the-shelf devices. The methodology developed here can be applied to high speed systems \cite{2015arXiv150908137C} and in a network setting \cite{2015arXiv150908389T}. In future, it will be intersting to consider other source flaws in MDI-QKD, for example, imperfect phase randomizations \cite{NJP.17.053014}. 

\begin{acknowledgements}
We thank F. Xu, E. Zhu,  and M. Curty for enlightening discussions. Financial support from NSERC Discovery Grant, NSERC RTI Grant, and the Canada Research Chairs Program is gratefully acknowledged.
\end{acknowledgements}

\input{./Supplemental_material/Expt_loss_tolerant_mdiqkd_supplemental}

\bibliography{qhe}



\end{document}

%% file: Supplemental_material/Expt_loss_tolerant_mdiqkd_supplemental.tex
\section{Supplemental Material}
\subsection{Upper-bounding phase error rate by rejected data analysis }
In this section, we give the algorithm used in the Letter to estimate the phase error rate $e^{11}_X$ using the rejected data analysis as proposed in \cite{PhysRevA.90.052314}.

In the actual three-state MDI-QKD protocol, Alice and Bob  send the untrusted third party Eve photons encoded in one of the three polarization states. Let  $|\phi_{j_\alpha}\rangle_{A_eE}$ be the purification of the state $\rho_{E,j_{\alpha}}$ sent by Alice to Eve, where $j_{\alpha}\in\{0_Z,1_Z,0_X\}$, and the subscripts $A_e$ and $E$ represent the extended system possessed by Alice and the system to be sent to Eve, respectively. Sending the state $\rho_{E,0_Z}$ ($\rho_{E,1_Z}$) to Eve by Alice is equivalent to preparing the tripartite state of systems $A$, $A_e$, $E$ 
\begin{equation}
\label{Alice_hybrid_state}
|\Psi\rangle_{AA_eE}=\frac{1}{\sqrt{2}}(|0_Z\rangle_A|\phi_{0_Z}\rangle_{A_eE}+|1_Z\rangle_A|\phi_{1_Z}\rangle_{A_eE})
\end{equation}
followed by a projective measurement on system $A$ in the $Z$ basis with an outcome of 0  (1), and sending system $E$ to Eve. 

Likewise, sending  $\rho_{0_Z}$ ($\rho_{1_Z}$) to Eve by Bob is equivalent to preparing the tripartite state $|\Psi\rangle_{BB_eE'}$ with systems $B$, $B_e$, and $E'$,
\begin{equation}
|\Psi\rangle_{BB_eE'}=\frac{1}{\sqrt{2}}(|0_Z\rangle_B|\phi_{0_Z}\rangle_{B_eE'}+|1_Z\rangle_B|\phi_{1_Z}\rangle_{B_eE'}),
\end{equation}
followed by a projective measurement on system $B$ in the $Z$ basis with outcome 0 (1), and sending system $E'$ to Eve. 

Now consider the following virtual protocol. Alice prepares the state $|\Psi\rangle_{AA_eE}$, measures system A in the X basis with outcome $j\in\{0,1\}$, and sends Eve the system E. The state sent to Eve can be written as
\begin{equation}
\label{Alice_vir_state}
\hat{\sigma}_{E, j_X}^{vir}=Tr_{AA_e}[\hat{P}(|j_X\rangle_A ) \mathbb{I}_{A_eE} \hat{P}(|\Psi\rangle_{AA_eE} )],
\end{equation}
where $\hat{P}(x)=|x\rangle\langle x|$, and $|j_X\rangle=1/\sqrt{2}(|0_Z\rangle+(-1)^j|1_Z\rangle)$.
Similarly, Bob prepares the state $|\Psi\rangle_{BB_eE}$,  measures system B in the $X$ basis  with outcome $s\in\{0,1\}$, and sends Eve the system $E'$ whose state is given by
\begin{equation} 
\hat{\sigma}_{E', s_X}^{vir}=Tr_{BB_e}[\hat{P}(|s_X\rangle_B ) \mathbb{I}_{B_eE'} \hat{P}(|\Psi\rangle_{BB_eE'} )],
\end{equation}
where $|s_X\rangle=1/\sqrt{2}(|0_Z\rangle+(-1)^s|1_Z\rangle)$.

The phase error rate of single photon components is determined by the transmission rates of the fictitious states:
\begin{equation}
\label{ex}
e_X^{11}=\frac{Y_{0_X1_X}^{\Psi^+,vir}+Y_{1_X0_X}^{\Psi^+,vir}}{Y_{0_X0_X}^{\Psi^+,vir}+Y_{1_X1_X}^{\Psi^+,vir} + Y_{0_X1_X}^{\Psi^+,vir}+Y_{1_X0_X}^{\Psi^+,vir}}.
\end{equation}
where 
 $Y_{j_Xs_X}^{\Psi^+}$ is the probability that Alice and Bob send Eve the virtual states $\hat{\sigma}_{E, j_X}$ and $\hat{\sigma}_{E', s_X}$, respectively, and Eve gets a successful Bell state measurement with outcome $|\Psi^+\rangle=(|H\rangle|V\rangle+|V\rangle|H\rangle)/\sqrt{2}$, which is given by

\begin{equation}
\label{Yvir0}
 Y_{j_Xs_X}^{\Psi^+,vir}=
 Tr[\hat{\sigma}_{E, j_X}^{vir}]
 Tr[\hat{\sigma}_{E', s_X}^{vir}]
 Tr[\hat{D}_{\Psi^+} \hat{\sigma'}_{E, j_X}^{vir}  \otimes \hat{\sigma'}_{E', s_X}^{vir}].
 \end{equation}

In the above equation, the operator $\hat{D}_{\psi^+}$ is Eve's operation corresponding to the BSM with outcome $\Psi^+$, and the operators $\hat{\sigma'}_{E, j_X}^{vir}$ and $\hat{\sigma'}_{E', j_X}^{vir}$ are the normalized versions of $\hat{\sigma}_{E, j_X}^{vir}$ and $\hat{\sigma}_{E', j_X}^{vir}$ given by
\begin{equation}
\begin{aligned}
&\hat{\sigma'}_{E, j_X}^{vir}=\hat{\sigma}_{E, j_X}^{vir}/Tr[\hat{\sigma}_{E, j_X}^{vir}]\\
&\hat{\sigma'}_{E', j_X}^{vir}=\hat{\sigma}_{E', j_X}^{vir}/Tr[\hat{\sigma}_{E', j_X}^{vir}]
\end{aligned}
\end{equation}

 The density operators of the virtual states $\hat{\sigma}_{E, j_X}^{vir}$ and $\hat{\sigma}_{E', s_X}^{vir}$ can be found from the density operators of the actual states $\rho_{j_\alpha}$. From Eqs. (\ref{Alice_hybrid_state}) and (\ref{Alice_vir_state}), the virtual state $\sigma_{E, j_X}$ sent to Eve by Alice is
\begin{equation}
\label{Alice_vir2}
\begin{aligned}
 \hat{\sigma}_{E, j_X}^{vir}
 =&
 Tr_{AA_e}[\hat{P}(|j_X\rangle_A ) \mathbb{I}_{A_eE} \hat{P}(|\Psi\rangle_{AA_eE} )]\\
 =&\frac{1}{4}[(\rho_{E,0_Z} + \rho_{E,1_Z})
 +(-1)^j 
 Tr_{A_e}(|\phi_{1_Z}\rangle_{A_eE} \langle \phi_{0_Z}|_{A_eE} \\
 &+ |\phi_{0_Z}\rangle_{A_eE} \langle \phi_{1_Z}|_{A_eE})
 ].
 \end{aligned}
 \end{equation}

Let $|\gamma_{j_\alpha}^0\rangle_E$ and $|\gamma_{j_\alpha}^1\rangle_E$ be the eigenvectors of $\rho_{E,j_\alpha}$, and $|\lambda_{E,j_\alpha}^0|^2$ and $|\lambda_{E,j_\alpha}^1|^2$ be the corresponding eigenvalues. The Schmidt decomposition of $|\phi_{j_\alpha}\rangle_{A_eE}$ is 
\begin{equation}
\label{Alice_schmidt}
|\phi_{j_\alpha}\rangle_{A_eE}
=
\lambda_{E,j_\alpha}^0
|0\rangle_{A_e}
|\gamma_{j_\alpha}^0\rangle_E
+
\lambda_{E,j_\alpha}^1
|1\rangle_{A_e}
|\gamma_{j_\alpha}^1\rangle_E
\end{equation}
where $\{|0\rangle_{A_e}\, |1\rangle_{A_e}\}$ is a basis of Alice's extended system $A_e$. Note that since Alice possesses the extended system $A_e$, she can select the basis $\{|0\rangle_{A_e}\, |1\rangle_{A_e}\}$ in the purification of $\rho_{E, j_{\alpha}}$ to optimize the key rate. In this paper, we use the same basis $\{|0\rangle_{A_e}\, |1\rangle_{A_e}\}$ for the purification of $\rho_{E,0_Z}$ and $\rho_{E,1_Z}$, which is not necessarily the optimal choice. The key rate can be further improved by optimizing the purification, which is left as future work.

Substituting Eq. (\ref{Alice_schmidt}) into (\ref{Alice_vir2}), the virtual state $\hat{\sigma}_{E, j_X}^{vir}$ is

\begin{equation}
\begin{aligned}
 \hat{\sigma}_{E, j_X}^{vir}
 & =
 \frac{1}{4}\{(\rho_{E,0_Z} + \rho_{E,1_Z})\\
 &+(-1)^j 
 [
 \lambda_{E,0_Z}^{0}\lambda_{E,1_Z}^{0}
 (
 |\gamma_{0_Z}^0\rangle_E \langle\gamma_{1_Z}^0|_E
 +
 |\gamma_{1_Z}^0\rangle_E \langle\gamma_{0_Z}^0|_E 
 )\\
 &+
 \lambda_{E,0_Z}^{1}\lambda_{E,1_Z}^{1}
 (
 |\gamma_{0_Z}^1\rangle_E \langle\gamma_{1_Z}^1|_E
 +
 |\gamma_{1_Z}^1\rangle_E \langle\gamma_{0_Z}^1|_E
 ) 
 ]
 \}.
 \end{aligned}
 \end{equation}

The density operator $\sigma_{E', s_X}^{vir}$ (the virtual state sent to Eve by Bob) can be found using the same method.

 We first discuss the case where the states lie in the $X-Z$ plane.  In this case, the Stokes parameter $S^Y=0$, and the states $\hat{\sigma'}_{E, j_X}^{vir}$ (with Stokes parameters ($S_{E,j_X}^{vir, X}$, $0$, $S_{E,j_X}^{vir,Z}$ ) and $\hat{\sigma'}_{E', s_X}^{vir}$  (with Stokes parameters ($S_{E,j_X}^{vir, X}$, $0$, $S_{E,j_X}^{vir,Z}$ ) can be written as a linear combination of the identity matrix $\hat{\sigma}_{I}$ and the Pauli matrices $\hat{\sigma}_{X}, \hat{\sigma}_{Z}$:
\begin{equation}
\label{sigmaE}
\hat{\sigma'}_{E,j_X}^{vir}
=
\dfrac{1}{2}
(
\hat{\sigma}_{I}
+
S_{E,j_X}^{vir, X}\hat{\sigma}_X
+
S_{E,j_X}^{vir,Z}\hat{\sigma}_Z
)
\end{equation} 

\begin{equation}
\label{sigmaE'}
\hat{\sigma'}_{E',s_X}^{vir}
=
\dfrac{1}{2}
(\hat{\sigma}_{I}
+
S_{E',s_X}^{vir, X}
\hat{\sigma}_X
+
S_{E',s_X}^{vir,Z}\hat{\sigma}_Z)
\end{equation} 

Define the transmission rate of $\hat{\sigma}_{t}  \otimes \hat{\sigma}_{t'}$, $t, t' \in\{I, X, Z\}$ as
\begin{equation}
\label{q}
q_{\Psi^+|t,t'}=\frac{1}{4}
Tr[\hat{D}_{\Psi^+} 
\hat{\sigma}_{t}  
\otimes 
\hat{\sigma}_{t'}].
\end{equation}

From Eqs. (\ref{Yvir0}) and (\ref{q}), the transmission rate $Y_{j_Xs_X}^{\Psi^+}$ can be written as 
\begin{widetext}
\begin{equation}
\label{Yvir}
\begin{aligned}
 Y_{j_Xs_X}^{\Psi^+,vir}
 &=
 Tr[\hat{\sigma}_{E, j_X}^{vir}]
 Tr[\hat{\sigma}_{E', s_X}^{vir}]
 \times 
( 
q_{\Psi^+| I \otimes I}
+
S_{E',s_X}^{vir, X} q_{\Psi^+|I \otimes X}
+
S_{E',s_X}^{vir, Z} q_{\Psi^+|I \otimes Z} \\
&+
S_{E,j_X}^{vir, X} q_{\Psi^+|X \otimes I}
+ 
S_{E,j_X}^{vir, X} S_{E',s_X}^{vir, X} q_{\Psi^+|X \otimes X}
+
S_{E,j_X}^{vir, X} S_{E',s_X}^{vir, Z} q_{\Psi^+|X \otimes Z} \\
&+
S_{E,j_X}^{vir, Z} q_{\Psi^+|Z \otimes I}
+ 
S_{E,j_X}^{vir, Z} S_{E',s_X}^{vir, X} q_{\Psi^+|Z \otimes X}
+
S_{E,j_X}^{vir, Z} S_{E',s_X}^{vir, Z} q_{\Psi^+|Z \otimes Z} 
)
\end{aligned}
\end{equation}
\end{widetext}

Let $\mathbf{S_{j_Xs_X}^{vir}}$ be a row vector and $\mathbf{q}$ be a column vector defined as 
\begin{widetext}

\begin{equation}
\mathbf{S_{j_Xs_X}^{vir}}
=
[
1, 
S_{E',s_X}^{vir, X}, \thinspace
S_{E',s_X}^{vir, Z}, \thinspace
S_{E,j_X}^{vir, X},  \thinspace
S_{E,j_X}^{vir, X} S_{E',s_X}^{vir, X},  \thinspace
S_{E,j_X}^{vir, X} S_{E',s_X}^{vir, Z},  \thinspace
S_{E,j_X}^{vir, Z},  \thinspace
S_{E,j_X}^{vir, Z} S_{E',s_X}^{vir, X},  \thinspace
S_{E,j_X}^{vir, Z} S_{E',s_X}^{vir, Z}  \thinspace
],
\end{equation}

\begin{equation}
\mathbf{q}
=
[
q_{\Psi^+|I \otimes I},  \thinspace
q_{\Psi^+|I \otimes X},  \thinspace
q_{\Psi^+|I \otimes Z},  \thinspace
q_{\Psi^+|X \otimes I},  \thinspace
q_{\Psi^+|X \otimes X},  \thinspace
q_{\Psi^+|X \otimes Z},  \thinspace
q_{\Psi^+|Z \otimes X},  \thinspace
q_{\Psi^+|Z \otimes Z}  \thinspace
]^T,
\end{equation}
\end{widetext}
respectively. The expression for the transmission rate $ Y_{j_Xs_X}^{\Psi^+,vir}$ (Eq. (\ref{Yvir})) can be written as 

\begin{equation}
\label{Yvir2}
Y_{j_Xs_X}^{\Psi^+,vir}
=
 Tr[\hat{\sigma}_{E, j_X}^{vir}]
 Tr[\hat{\sigma}_{E', s_X}^{vir}]
 \mathbf{S_{j_Xs_X}^{vir}} \mathbf{q}.
\end{equation}
Once we know the transmission rates of the Pauli matrices $\mathbf{q}$, we can estimate $Y_{j_Xs_X}^{\Psi^+,vir}$ and the phase error rate $e_X^{11}$. In the next session, we will discuss how to find $\mathbf{q}$ from experimental data.

When the states $\rho_{E, j_{\alpha}}$  prepared by Alice / Bob do not lie in the $X-Z$ plane, we can always find a reference frame such that the states $\rho_{E,0_Z}$, $\rho_{E,1_Z}$, and $\rho_{E,0_X}$ have a common Stokes parameter $S^Y_E$ (i.e., the Stokes parameters of the state $\rho_{E, j_{\alpha}}$ is given by ($S_{E, j_{\alpha}}^X$, $S^Y_E $, $S_{E, j_{\alpha}}^Z$).) We apply the filtering technique described in \cite{PhysRevA.90.052314}, which shows that, for a state $\rho_{E, j_{\alpha}}$ with a nonzero $S_E^Y$, we can equivalently consider the following state with its Stokes parameters given by,
\begin{equation}
(\frac{S_{E, j_{\alpha}}^X}{f(q)}, 0, \frac{S_{E, j_{\alpha}}^Z}{f(q)})
\end{equation}
where $f(q)$ is given by
\begin{equation}
f(q)=\frac{2(1-q)q}{1-2q+2q^2}
\end{equation}
and $q$ is determined by solving the following equation
\begin{equation}
S_E^Y =  \frac{(2q-1)}{(1-2q+2q^2)}.
\end{equation}

\subsubsection{Estimating transmission rates of Pauli matrices from experimental data}
In this section, we will show how to estimate the transmission rates of Pauli matrices $\mathbf{q}$ from experimental data. Recall in the three-state MDI-QKD, Alice (Bob) randomly sends Eve one of the three states $\rho_{E, 0_Z}$ ($\rho_{E', 0_Z}$), $\rho_{E, 1_Z}$ ($\rho_{E', 1_Z}$), $\rho_{E, 0_X}$ ($\rho_{E', 0_X}$). As in the previous section, the subscripts $E$ and $E'$ represent the systems sent to Eve by Alice and Bob, respectively.

 Let $Y_{j_\alpha s_\beta}^{\Psi^+,11}$ be the \textit{conditional} probability that Eve gets a successful Bell state measurement with outcome $\Psi^+$ given that Alice sends Eve a single photon of state $\rho_{E, j_\alpha}$ and Bob sends Eve a single photon of state $\rho_{E', s_\beta}$ (the superscript 11 represent that both Alice and Bob send out single photons). Following the procedures described in the previous section, the transmission rate of the actual states  $Y_{j_\alpha s_\beta}^{\Psi^+,11}$ can be written as

\begin{equation}
 Y_{j_\alpha s_\beta}^{\Psi^+,11}
 =
 \mathbf{S_{j_{\alpha}s_{\beta}}} 
 \mathbf{q},
\end{equation}
where $\mathbf{S_{j_{\alpha}s_{\beta}}} $ is related to the actual states  $\rho_{E, j_{\alpha}}$  (with Stokes parameters ($S_{E,j_\alpha}^{X}$, $0$, $S_{E,j_\alpha}^{Z}$) and  $\rho_{E', s_{\beta}}$  (with Stokes parameters ($S_{E',s_beta}^{X}$, $0$, $S_{E',s_\beta}^{Z}$) as follows:

\begin{widetext}
\begin{equation}
\mathbf{S_{j_{\alpha}s_{\beta}}}
=
[
1, 
S_{E',s_\beta}^{ X}, \thinspace
S_{E',s_\beta}^{Z}, \thinspace
S_{E,j_\alpha}^{X},  \thinspace
S_{E,j_\alpha}^{X} S_{E',s_\beta}^{X},  \thinspace
S_{E,j_\alpha}^{X} S_{E',s_\beta}^{Z},  \thinspace
S_{E,j_\alpha}^{Z},  \thinspace
S_{E,j_\alpha}^{Z} S_{E',s_\beta}^{X},  \thinspace
S_{E,j_\alpha}^{Z} S_{E',s_\beta}^{Z}  \thinspace
].
\end{equation}
\end{widetext}

From experiment, we can get the following set of independent linear equations:

\begin{equation}
\label{linsys}
\begin{aligned}
&Y_{0_Z 0_Z}^{\Psi^+,11} = \mathbf{S_{0_{Z}0_{Z}}} \mathbf{q}, \\
&Y_{0_Z 1_Z}^{\Psi^+,11} = \mathbf{S_{0_{Z}1_{Z}}} \mathbf{q}, \\
&Y_{1_Z 0_Z}^{\Psi^+,11} = \mathbf{S_{1_{Z}0_{Z}}} \mathbf{q}, \\
&Y_{1_Z 1_Z}^{\Psi^+,11} = \mathbf{S_{1_{Z}1_{Z}}} \mathbf{q}, \\
&Y_{0_X 0_Z}^{\Psi^+,11} = \mathbf{S_{0_{X}0_{Z}}} \mathbf{q}, \\
&Y_{0_X 1_Z}^{\Psi^+,11} = \mathbf{S_{0_{X}1_{Z}}} \mathbf{q}, \\
&Y_{0_Z 0_X}^{\Psi^+,11} = \mathbf{S_{0_{Z}0_{X}}} \mathbf{q}, \\
&Y_{1_Z 0_X}^{\Psi^+,11} = \mathbf{S_{1_{Z}0_{X}}} \mathbf{q}, \\
&Y_{0_X 0_X}^{\Psi^+,11} = \mathbf{S_{0_{X}0_{X}}} \mathbf{q}. \\
\end{aligned}
\end{equation}

Define a vector $\mathbf{Y^{\Psi^+,11}}$

\begin{equation}
\begin{aligned}
\mathbf{Y^{\Psi^+,11}}
=[
Y_{0_Z 0_Z}^{\Psi^+,11},  \thinspace
Y_{0_Z 1_Z}^{\Psi^+,11},  \thinspace 
Y_{1_Z 0_Z}^{\Psi^+,11},  \thinspace 
Y_{1_Z 1_Z}^{\Psi^+,11},  \thinspace \\
Y_{0_X 0_Z}^{\Psi^+,11},  \thinspace  
Y_{0_X 1_Z}^{\Psi^+,11},  \thinspace  
Y_{0_Z 0_X}^{\Psi^+,11},  \thinspace  
Y_{1_Z 0_X}^{\Psi^+,11},  \thinspace  
Y_{0_X 0_X}^{\Psi^+,11}  \thinspace  
]
\end{aligned}
\end{equation}

and a matrix $\mathbbm{S}$
\[ 
\mathbbm{S}=
\begin{bmatrix}
\mathbf{S_{0_{Z}0_{Z}}} \\
\mathbf{S_{0_{Z}1_{Z}}} \\
\mathbf{S_{1_{Z}0_{Z}}} \\
\mathbf{S_{1_{Z}1_{Z}}} \\
\mathbf{S_{0_{X}0_{Z}}} \\
\mathbf{S_{0_{X}1_{Z}}} \\ 
\mathbf{S_{0_{Z}0_{X}}} \\
\mathbf{S_{1_{Z}0_{X}}} \\
\mathbf{S_{0_{X}0_{X}}} \\
\end{bmatrix}
\]

The linear system (\ref{linsys}) can be concisely written as 
\begin{equation}
\label{linsys2}
\mathbf{Y^{\Psi^+,11}}=\mathbbm{S}\mathbf{q}.
\end{equation}

Knowing $\mathbf{Y^{\Psi^+,11}}$ from the experiment, the transmission rates $\mathbf{q}$ can be solved:
\begin{equation}
\mathbf{q}=\mathbbm{S}^{-1}\mathbf{Y^{\Psi^+,11}}.
\end{equation}

The transmission rates of the virtual states can then be calculated by Eq.(\ref{Yvir2}), and the phase error rate can be estimated by Eq.(\ref{ex}).

\subsubsection{Bounding $e_X^{11}$ with a finite number of decoy states}
In the previous two sections, we give the method to estimate the phase error rate $e_X^{11}$ from the $Y_{j_\alpha s_\beta}^{\Psi^+,11}$, which is the yield of single photon components. The parameter $Y_{j_\alpha s_\beta}^{\Psi^+,11}$ can be precisely estimated with an infinite number of decoy states. 

In reality, we can only apply a finite number of decoy states, where the value of $Y_{j_\alpha s_\beta}^{\Psi^+,11}$ can not be precisely determined. Instead, we can find an upper bound $Y_{j_\alpha s_\beta}^{\Psi^+,11,U}$, and a lower bound $Y_{j_\alpha s_\beta}^{\Psi^+,11,L}$, either analytically \cite{NJP.15.113007, NatureCommun.5.3732} or by linear programming. In this case, the linear system (\ref{linsys}, \ref{linsys2}) should be replaced with the following linear inequality:
\begin{equation}
\label{Yconstraint}
\mathbf{Y^{\Psi^+,11,L}}
\leq
\mathbbm{S}\mathbf{q}
\leq
\mathbf{Y^{\Psi^+,11,U}}.
\end{equation}
where 

\begin{equation}
\begin{aligned}
\mathbf{Y^{\Psi^+,11,L}}
=[
Y_{0_Z 0_Z}^{\Psi^+,11,L},  \thinspace
Y_{0_Z 1_Z}^{\Psi^+,11,L},  \thinspace 
Y_{1_Z 0_Z}^{\Psi^+,11,L},  \thinspace 
Y_{1_Z 1_Z}^{\Psi^+,11,L},  \thinspace \\
Y_{0_X 0_Z}^{\Psi^+,11,L},  \thinspace  
Y_{0_X 1_Z}^{\Psi^+,11,L},  \thinspace  
Y_{0_Z 0_X}^{\Psi^+,11,L},  \thinspace  
Y_{1_Z 0_X}^{\Psi^+,11,L},  \thinspace  
Y_{0_X 0_X}^{\Psi^+,11,L}  \thinspace  
],
\end{aligned}
\end{equation}

and

\begin{equation}
\begin{aligned}
\mathbf{Y^{\Psi^+,11,U}}
=[
Y_{0_Z 0_Z}^{\Psi^+,11,U},  \thinspace
Y_{0_Z 1_Z}^{\Psi^+,11,U},  \thinspace 
Y_{1_Z 0_Z}^{\Psi^+,11,U},  \thinspace 
Y_{1_Z 1_Z}^{\Psi^+,11,U},  \thinspace \\
Y_{0_X 0_Z}^{\Psi^+,11,U},  \thinspace  
Y_{0_X 1_Z}^{\Psi^+,11,U},  \thinspace  
Y_{0_Z 0_X}^{\Psi^+,11,U},  \thinspace  
Y_{1_Z 0_X}^{\Psi^+,11,U},  \thinspace  
Y_{0_X 0_X}^{\Psi^+,11,U}  \thinspace  
].
\end{aligned}
\end{equation}

Our task is to find an upper bound of the phase error rate $e_X^{11}$. The expression of $e_X^{11}$ (Eq. \ref{ex}) can be rewritten as
\begin{equation}
e_X^{11}=\frac{1}{1 + \frac{Y_{0_X0_X}^{\Psi^+,vir}+Y_{1_X1_X}^{\Psi^+,vir}}{Y_{0_X1_X}^{\Psi^+,vir}+Y_{1_X0_X}^{\Psi^+,vir}}}.
\end{equation}
The upper bound of $e_X^{11}$ found by lower bounding $Y_{0_X0_X}^{\Psi^+,vir}+Y_{1_X1_X}^{\Psi^+,vir}$ and upper bounding $Y_{0_X1_X}^{\Psi^+,vir}+Y_{1_X0_X}^{\Psi^+,vir}$:
\begin{equation}
e_X^{11}
\leq
e_X^{11,U}
=
\frac{1}{1 + \frac{(Y_{0_X0_X}^{\Psi^+,vir}+Y_{1_X1_X}^{\Psi^+,vir})^L}{(Y_{0_X1_X}^{\Psi^+,vir}+Y_{1_X0_X}^{\Psi^+,vir})^U } }.
\end{equation}

Finding a lower bound of  $Y_{0_X0_X}^{\Psi^+,vir}+Y_{1_X1_X}^{\Psi^+,vir}$ is equivalent to the following linear programming problem:

\begin{equation}
\label{CorrL}
\begin{aligned}
\underset{q}{\min}
&\{
(
Tr[\hat{\sigma}_{E, 0_X}^{vir}]
 Tr[\hat{\sigma}_{E', 0_X}^{vir}]
  \mathbf{S_{0_X0_X}^{vir}}  \\
&+
Tr[\hat{\sigma}_{E, 1_X}^{vir}]
 Tr[\hat{\sigma}_{E', 1_X}^{vir}]
  \mathbf{S_{1_X1_X}^{vir}} 
 )
\mathbf{q}
\}
 \end{aligned}
\end{equation}
subject to the constraint given by inequality (\ref{Yconstraint}). 

Similarly, upper bounding  $Y_{0_X1_X}^{\Psi^+,vir}+Y_{1_X0_X}^{\Psi^+,vir}$ is equivalent to the following linear programming problem:

\begin{equation}
\label{ErroU}
\begin{aligned}
\underset{q}{\max}
&\{
(
Tr[\hat{\sigma}_{E, 0_X}^{vir}]
 Tr[\hat{\sigma}_{E', 1_X}^{vir}]
  \mathbf{S_{0_X1_X}^{vir}}  \\
&+
Tr[\hat{\sigma}_{E, 1_X}^{vir}]
 Tr[\hat{\sigma}_{E', 0_X}^{vir}]
  \mathbf{S_{1_X0_X}^{vir}} 
 )
\mathbf{q}
\}
 \end{aligned}
\end{equation}
subject to the constraint (\ref{Yconstraint}).

\subsection{State Characterization}
In this session, we discuss the sources of errors involved in preparing the BB84 states, and how we minimize the state preparation errors. We then present the details of state characterization using quantum state tomography.

\subsubsection{Sources of encoding errors}
\begin{figure}[b]
  \centering
  \includegraphics[width=0.3\textwidth]{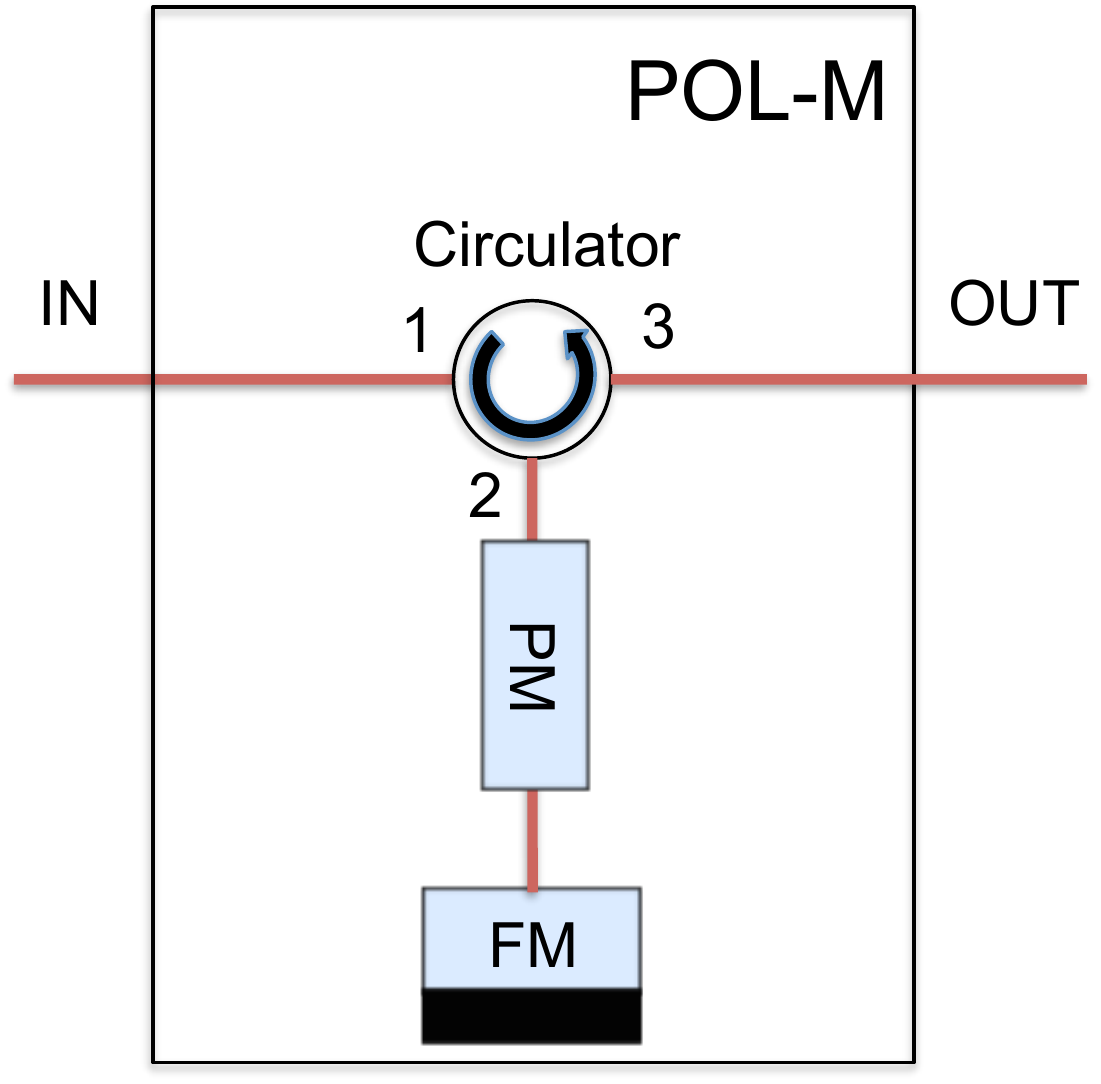}
  \caption{\label{polm}(Color online). Schematic of the polarization modulator.}
 \end{figure}

Fig. \ref{polm} shows the schematic of the bi-directional polarization modulator \cite{NJP.11.095001}. Optical pulses are launched through an optical circulator to a phase modulator (PM). The polarization of the light is at $45^{\circ}$ to the TE axis of the $\mathrm{LiNbO_3}$ waveguide inside the PM. When an optical pulse travels through the PM for the first time, a positive voltage $+V$ is applied on the phase modulator. The pulse is reflected by a Faraday mirror (FM) with its polarization rotated by $90^{\circ}$, and travels back. When the pulse travels through the PM for the second time, a negative voltage $-V$ is applied on the PM. Due to the different modulation efficiency in the TE and TM modes, we introduce a phase difference along the TE and TM directions. The output state can be expressed as 
\begin{equation}
\label{outputState}
|\psi\rangle=\frac{|TE\rangle+e^{i\psi}|TM\rangle}{\sqrt{2}},
\end{equation}
where $|TE\rangle$ and $|TM\rangle$ represent the polarization states along the TE and TM directions of the PM's waveguide, and $\psi$ is the phase difference introduced, which depends on the applied voltage. By modulating $\psi$ to \{$0$, $\pi$, $\pi/2$\}, we can generate the three states \{$\rho_{0_Z}$, $\rho_{1_Z}$, $\rho_{0_X}$\} needed in our protocol.

Here we discuss the sources of errors in the encoding system that lead to imperfect state preparations.

\textit{Power mismatch in TE and TM modes} Ideally we want optical pulses to be launched into the PM at an angle of $45^{\circ}$ relative to the TE axis of the PM's waveguide. Is this case, the powers along the TE and TM directions are equal, and the output states \{$\rho_{0_Z}$, $\rho_{1_Z}$, $\rho_{0_X}$\} are located on a great circle on the Poincar\'e sphere. However, optical pulses may be launched at an angle $\kappa$ other than $45^{\circ}$. In this case, the modulated output state \ref{outputState} should be rewritten as
\begin{equation}
|\psi\rangle=cos(\kappa)|TE\rangle+sin(\kappa)e^{i\psi}|TM\rangle.
\end{equation}

As a result, the output states \{$\rho_{0_Z}$, $\rho_{1_Z}$, $\rho_{0_X}$\} are distributed on a small circle on the Poincar\'e sphere. In this case, $\rho_{0_Z}$ and $\rho_{1_Z}$ are no longer orthogonal, and their overlap (characterized by $F(\rho_{0_Z}, \rho_{0_Z})^2$, where $F(\rho_{0_Z}, \rho_{0_Z})$ is the fidelity between $\rho_{0_Z}$ and $\rho_{1_Z}$) is $cos^2(2\kappa)$. This is the dominant error that lead to modulation errors in our encoding system. 

\textit{Control voltage accuracy}  The accuracy is limited by the voltage resolution of the signal source driving the PM. In our experiment, the waveform generator driving the PM has an output amplitude of $\pm$5 V and a resolution of 1 mV. The $V_{\pi}$ of the PM is around 5 V, which means that error due to limited resolution of the driving voltage is relatively small.

\begin{figure}[b]
  \centering
  \includegraphics[width=0.5\textwidth]{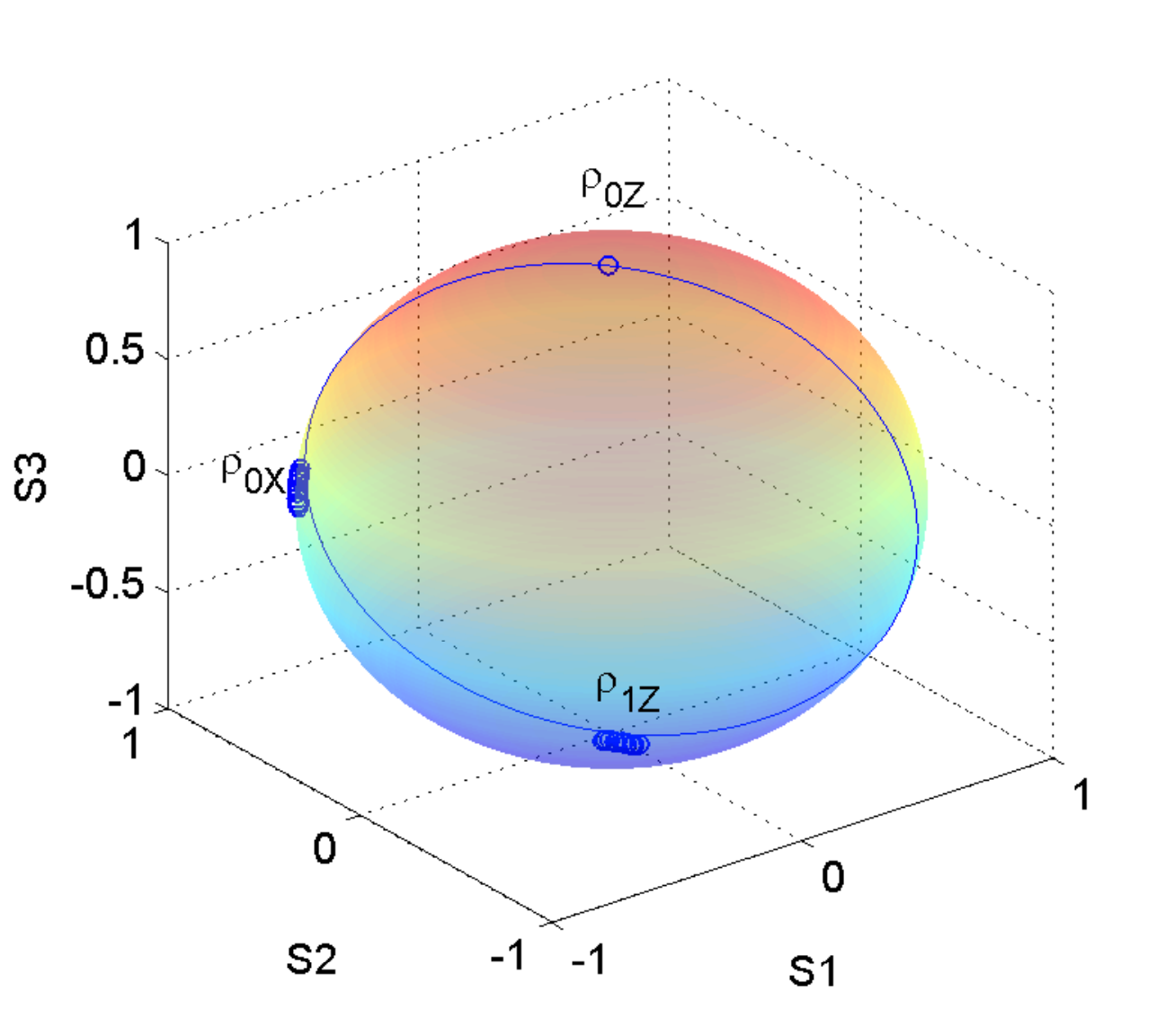}
  \caption{\label{scanStates}(Color online). Search of the states with minimum encoding errors. We scan the voltages applied on the polarization modulator to find the states $\rho_{0_X}$ and $\rho_{1_Z}$ with minimum modulation errors.}
 \end{figure}

\begin{figure}[t]
  \centering
  \includegraphics[width=0.5\textwidth]{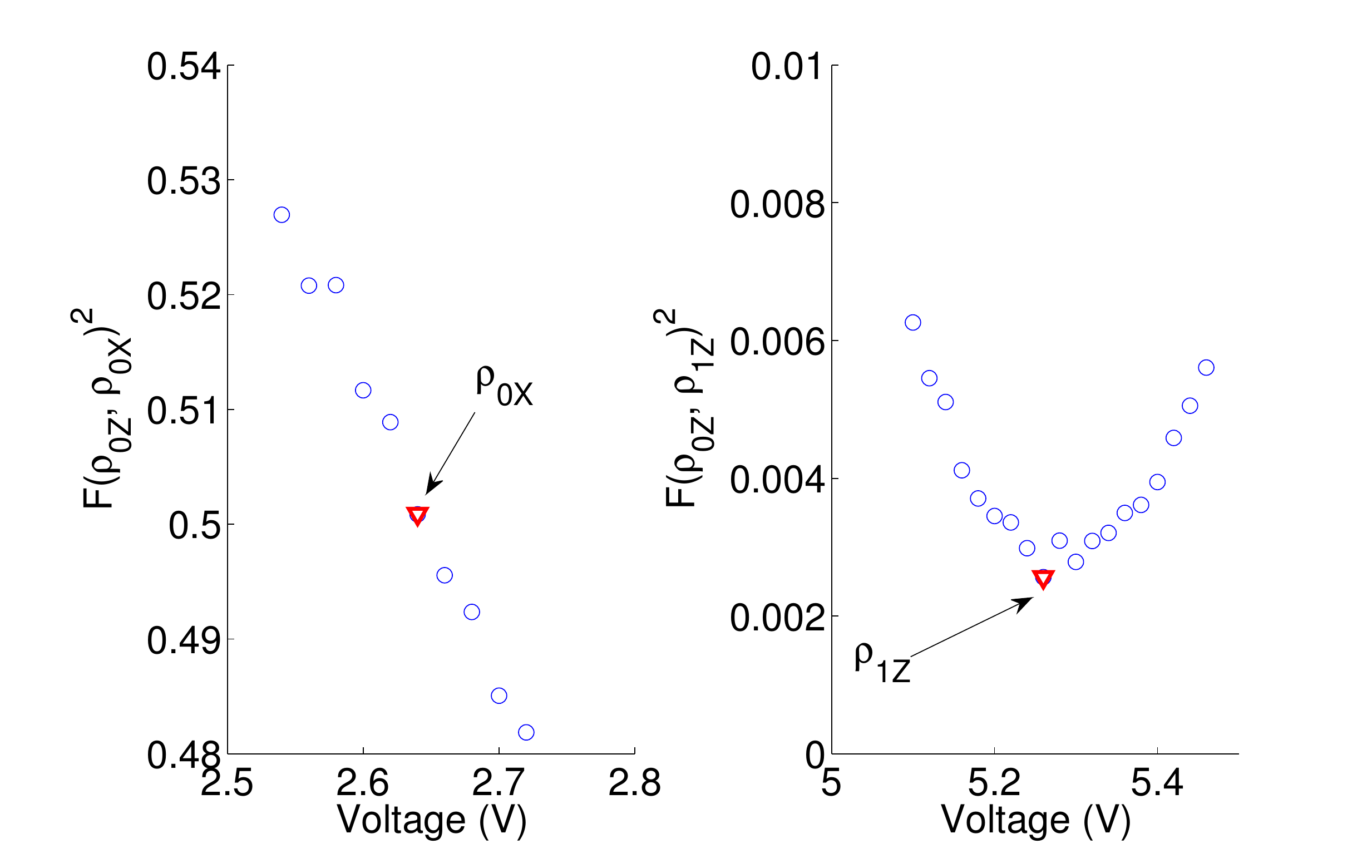}
  \caption{\label{scanStateFidel}(Color online). Overlap between $\rho_{0_Z}$ and $\rho_{0_X}$, and overlap between $\rho_{0_Z}$ and $\rho_{1_Z}$, with different voltages applied on the Pol-M.  The voltage for $\rho_{0_X}$ is chosen to get the overlap  between $\rho_{0_Z}$ and $\rho_{0_X}$ as close to 0.5 as possible, and the voltage for $\rho_{1_Z}$ is chosen such that the overlap between $\rho_{0_Z}$ and $\rho_{1_Z}$ is minimized.}
 \end{figure}

To minimize the errors in the state preparation, we finely scan the voltage applied on the phase modulator at a step of 0.02 V and characterize the corresponding output states. The step size of 0.02 V guarantees that the error due to voltage accuracy is less than 0.4\%. Fig. \ref{scanStates} shows different states corresponding to different voltages applied on the polarization modulator.  $\rho_{0_Z}$ corresponds to the state when the applied voltage is 0 V. We search around 2.5 V and 5.0 V at a step size of 0.02 V to search the states $\rho_{0_X}$ and $\rho_{1_Z}$ with minimum encoding errors. Each point on the Poincar\'e sphere corresponds one applied voltage. The states are reconstructed using quantum state tomography, as discussed in the next section. Fig. \ref{scanStateFidel} shows the overlap between $\rho_{0_Z}$ and $\rho_{0_X}$, and the overlap between $\rho_{0_Z}$ and $\rho_{1_Z}$, with different voltages. The voltage for $\rho_{0_X}$ is chosen such that the overlap  between $\rho_{0_Z}$ and $\rho_{0_X}$ is as close to 0.5 as possible, and the voltage for $\rho_{1_Z}$ is chosen such that the overlap between $\rho_{0_Z}$ and $\rho_{1_Z}$ is minimized.

\subsubsection{Quantum state tomography}

\begin{figure}[b]
  \centering
  \includegraphics[width=0.5\textwidth]{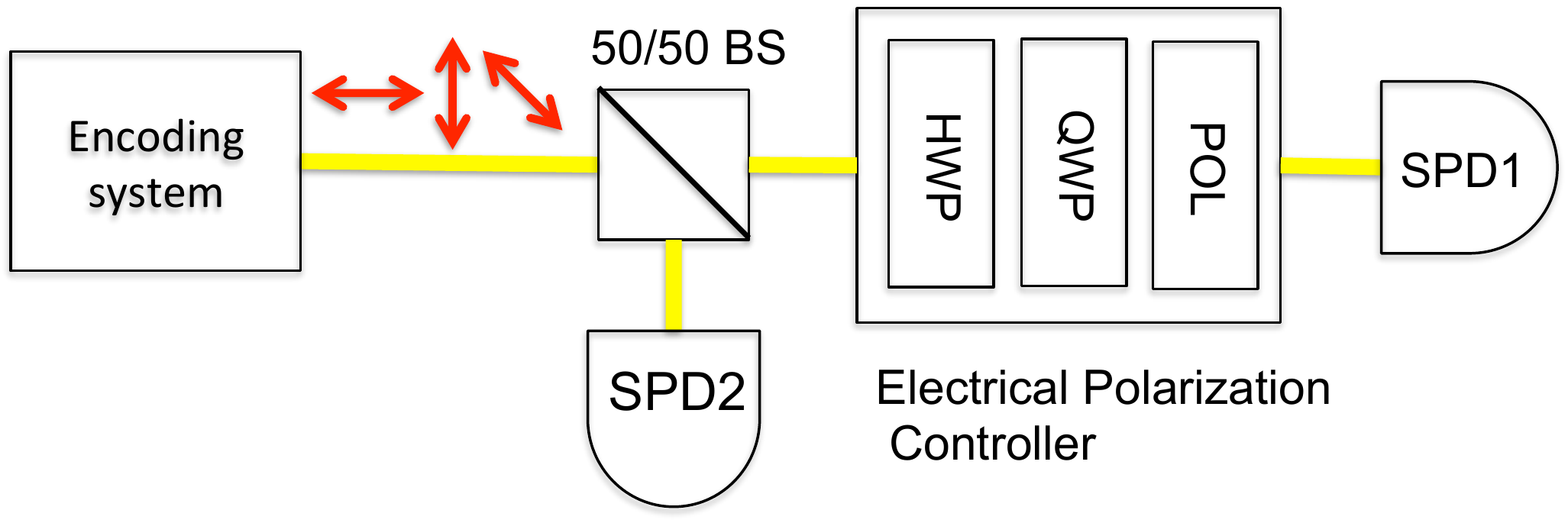}
  \caption{\label{tomography_setup}(Color online). Schematic of the quantum state tomography setup. HWP: half wave plate; QWP, quarter wave plate; POL: polarizer; SPD, single photon detector.}
 \end{figure}

Fig.\ref{tomography_setup} shows the setup of the quantum state tomography experiment. Optical pulses encoded in the polarization state $\rho_{j\alpha},$ where $j_{\alpha} \in \{0_Z, 1_Z, 0_X\}$,  are sent to the electrical polarization controller for projective measurements. The projective state $|\psi\rangle$ is given by
\begin{equation}
|\psi \rangle = U_{HWP}^{\dagger}(\theta) U_{QWP}^{\dagger}(\phi)  |H\rangle.
\end{equation}
The operations $U_{HWP}(\phi)$ and $U_{QWP}(\phi)$ are the unitary transformations by a half wave plate (HWP) and a quarter wave plate (QWP) with fast axes set to $\theta$ and $\phi$, respectively, which are given by
\[ 
U_{HWP}(\theta)=
\begin{bmatrix}
cos(2\theta) & sin(2\theta)\\
sin(2\theta) & -cos(2\theta) \\
\end{bmatrix}
\]
\[ 
U_{QWP}(\phi)=
\begin{bmatrix}
cos^2(\phi) + isin^2(\phi) & (1-i)cos(\phi)sin(\phi)\\
(1-i)cos(\phi)sin(\phi) & sin^2(\phi) + icos^2(\phi) \\
\end{bmatrix}
\].

In the tomography experiment, each state $\rho_{j_\alpha}$ is projected into the following four polarization basis states: horizontal $|H\rangle$, vertical $|V\rangle$, diagonal$|D\rangle$, and right-hand circular $|R\rangle$. The settings of the HWP, QWP, and POL are summarized in Table \ref{tomography_setting}. Photons are detected by a single photon detector (SPD1). Another single photon detector (SPD2) is used to monitor the total intensity of the incoming light pulses. The data acquisition time for each projective measurement is $t=10s$, and the counts are summarized in Table \ref{tomography_data1}.

 \begin{table}[b]
\caption{\label{tomography_setting} Angles of waveplates and polarizer angles for quantum state tomography.}
\begin{ruledtabular}
\begin{tabular}{ccccc}
Projective state & HWP & QWP & POL\\
  \hline
  $|H\rangle$ &  $0^\circ$ & $0^\circ$ & $0^\circ$\\
  $|V\rangle$ &  $45^\circ$ & $0^\circ$ & $0^\circ$\\
  $|D\rangle$ & $22.5^\circ$ & $0^\circ$ & $0^\circ$\\
  $|R\rangle$ &  $0^\circ$ & $45^\circ$ & $0^\circ$\\
 
\end{tabular}
\end{ruledtabular}
\end{table}

\begin{table}[b]
\caption{\label{tomography_data1} Raw counts in the quantum state tomography experiment. Counts are accumulated for 10 s.}
\begin{ruledtabular}
\begin{tabular}{ccccc}
\multirow{2}{*}{State} & \multicolumn{4}{c}{Projected states}  \\
  & $|H\rangle$ & $|V\rangle$ & $|D\rangle$ & $|R\rangle$\\
  \hline
  $\rho_{0_Z}$ & 201311 & 583 & 112867 & 114043\\
  $\rho_{1_Z}$ & 982 & 203500  & 122028 & 110687\\
  $\rho_{0_X}$ & 114815	 & 117459 & 35646 & 38239\\
 
\end{tabular}
\end{ruledtabular}
\end{table}

Below we describe the procedures to reconstruct the density matrices from the data in Table \ref{tomography_data1} (see next section) using the maximum likelihood technique \cite{QIC.3.503}. For each projective measurement, counts detected by SPD1 are accumulated for $10s$, and the results are shown in Table I of the main text. The total counts corresponding to the projective measurement to $|H\rangle$, $|V\rangle$, $|D\rangle$, and $|R\rangle$ are denoted as $n_H$, $n_V$, $n_D$, and $n_R$, respectively.  We first calculate a normalized count rate $\tilde{n}_{\psi}, \psi\in\{H, V, D, R\}$ to correct the impacts of dark counts and deadtime:
\begin{equation}
\tilde{n}_{\psi}
=
\frac{n_{\psi}}{t-n_{\psi}\tau} - DC
\end{equation}
where $t = 10s$ is the data acquisition time, $\tau=10 \mu s$  is the detector deadtime, and $DC = 50Hz$ is the dark count rate. Note that in the above expression, the term ($t-n_{\psi}\tau$) gives the total active time of the detector during $t$, and $\frac{n_{\psi}}{t-n_{\psi}\tau}$ gives the counting rate per unit active time.

The density matrix to be reconstructed can be written as 
\begin{equation}
\rho_{j_\alpha}=\frac{T_{j_\alpha}^{\dagger}T_{j_\alpha}}
{Tr[T_{j_\alpha}^{\dagger}T_{j_\alpha}]}.
\end{equation}
where $T^{\dagger}_{j_\alpha}$ is the conjugate transpose of $T_{j_\alpha}$, and $T_{j_\alpha}$ is given by

\[ 
T_{j_\alpha}=
\begin{bmatrix}
t_1 & 0\\
t_3+it_4 & t_2 \\
\end{bmatrix}
\].

The values of $t_1$, $t_2$, $t_3$, and $t_4$ are determined numerically by minimizing the following likelihood function:
\begin{equation}
L(t_1, t_2, t_3, t_4)=\sum_{\psi = H, V, D, R}
\frac{[N \langle \psi|\rho_{j_\alpha}(t_1, t_2, t_3, t_4) |\psi\rangle - \tilde{n}_{\psi} ]^2}
{2N\langle \psi|\rho_{j_\alpha}(t_1, t_2, t_3, t_4) |\psi\rangle}
\end{equation}
where $N = \tilde{n}_H + \tilde{n}_V$.

To estimate the error distributions, we use Monte Carlo simulations  to numerically generate additional data based on the experimental data and errors in the setup. As discussed in the Letter, the intensity and input polarization states are relatively stable and no drifts are observed within the span of the tomography measurement. See Fig. and its caption for details. 

\begin{figure}
  \centering
  \includegraphics[width=0.5\textwidth]{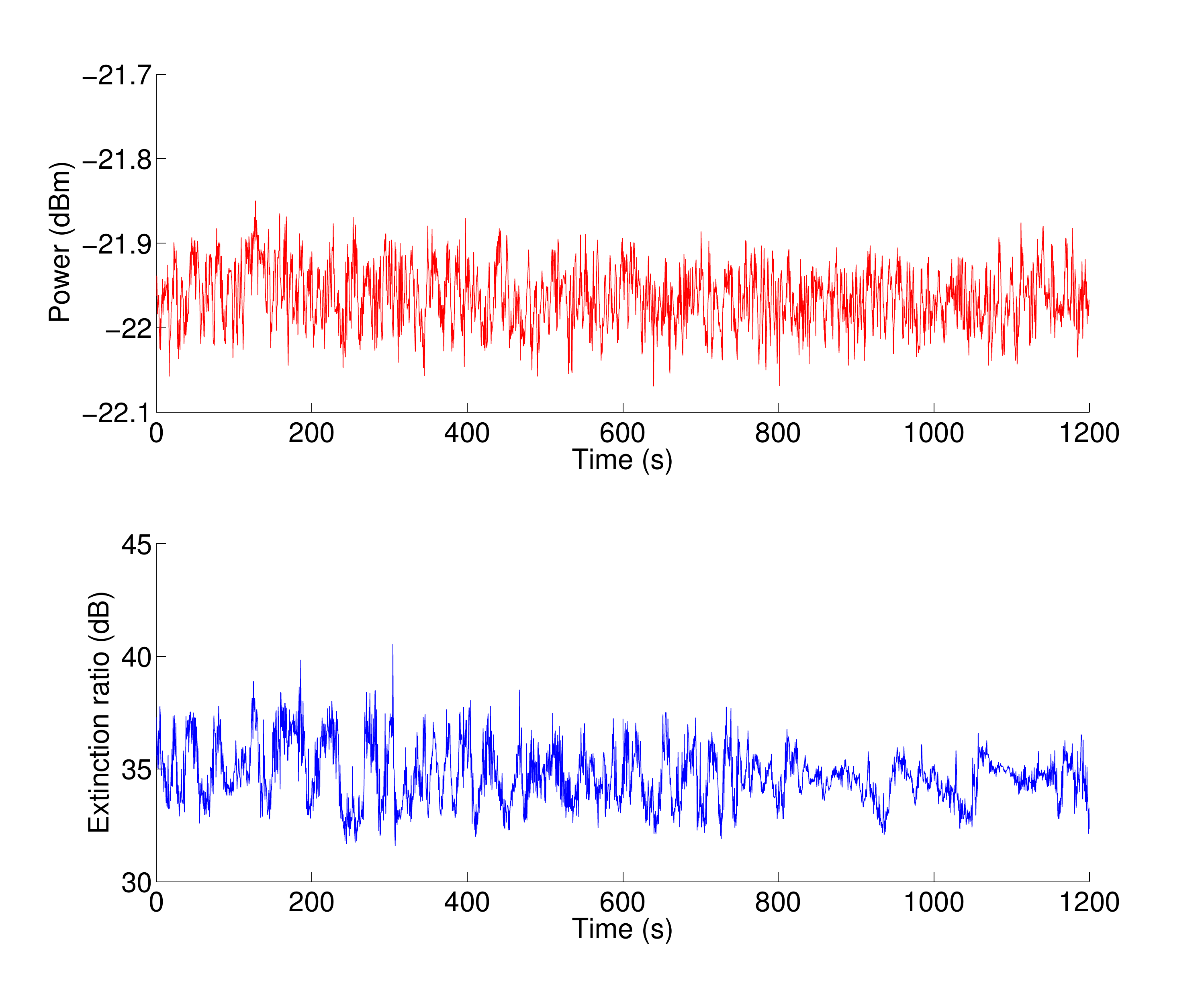}
  \caption{\label{state stability}(Color online). Stability of intensities and input states during the quantum state tomography. The upper figure shows the optical power coming out of the encoding system (intensity was not attenuated to single photon level in this measurement). The lower figure shows the stability test of the input polarization states. Horizontally polarized light coming out of the encoding system is measured at the H/V basis by a polarizing beam splitter. The figure shows the extinction ratio (i.e., the ratio of the power at the H and V output ports), which is around 35 dB over a period of 1200 s. The results show that the intensities and polarization states remain stable within the span of the tomography experiment ($<$ 5 minutes).  }
 \end{figure}

 We therefore consider two sources of errors: errors in counting statistics and errors in the angles of waveplates.  To simulate the errors in $n_\psi$, we assume the detection counts follow the Poisson distribution. In the simulation, a random number $n_{\psi}^{sim}$ is generated from the Poisson distribution with mean given by the experimental value $n_\psi$ as an input to the maximum likelihood algorithm.

Errors in setting waveplates' angles lead to projection to a state other than the one intended. Our electrical polarization controller has a accuracy of $\pm0.1^{\circ}$ ($0.00175$ rad)  in waveplate angle settings. To model errors in waveplate angles $\theta$ and $\phi$, random variables $\theta^{sim}$ and $\phi^{sim}$ are generated from the normal distributions $N(\theta, 0.00175^2)$ and $N(\phi, 0.00175^2)$, respectively, where $\theta$ and $\phi$ are the intended angle settings, and $N(\bar{(x}, \sigma^2)$ is the Gaussian distribution with mean $x$ and variance $\sigma^2$. The state projected into is given by $|\psi^{sim} \rangle = U_{HWP}^{\dagger}(\theta^{sim}) U_{QWP}^{\dagger}(\phi^{sim})  |H\rangle$, where $|H\rangle$ is the horizontal state given by $|H\rangle = [1, 0]^T$.

In each round of simulation, a set of data, including $n_{\psi}^{sim}$, $\theta^{sim}$, and $\phi^{sim}$ , are numerically generated using the distributions described above, and are used to calculate a density matrix using the maximum likelihood method. For each state $\rho_{j\alpha}$, a total of $1\times10^3$ sets of data are simulated to give the error distribution of the density matrix constructed.

\subsection{Experimental results}
We preformed the loss tolerant MDI-QKD experiment over 10 km and 40 km of optical fibers. The detailed experimental data is presented below.

\subsubsection{10 km loss tolerant MDI-QKD}
In this section, we present detailed experimental results not covered in the main text.

In the 10 km demonstration, we send a total of $6\times10^{11}$ pulses. The probabilities of sending $\rho_{0_Z}$, $\rho_{1_Z}$, and $\rho_{0_X}$ are  $P_{0_Z}=0.25$, $P_{1_Z}=0.25$, and $P_{0_X}=0.5$. The intensities of the signal state is $\mu = 0.2$ photon per pulse,  and the intensities of the decoy states are $\nu_1 = 0.03$ and $\nu_2 = 0$ photon per pulse.

Table \ref{Q} shows $Q_{j{_\alpha} s_{\beta}} ^{\Psi^+,I_A I_B}$, the conditional probability that Eve gets a successful Bell state measurement with outcome $\Psi^+$ given that Alice sends out a pulse of intensity $I_A$ in the state $\rho_{j_{\alpha}}$ and Bob sends out a pulse of intensity $I_B$ in the state $\rho_{s_\beta}$.

The upper and lower bounds of the yields of single photon components $Y_{j{_\alpha} s_{\beta}} ^{\Psi^+,11}$ are estimated given the following constraint \cite{PhysRevA.86.052305}:
\begin{equation}
\label{decoyConstraint}
\begin{aligned}
&Q_{j_{\alpha} s_{\beta}} ^{\Psi^+,I_A I_B}
(1 - \frac{k}{\sqrt{ N_{j_{\alpha} s_{\beta}} ^{I_A I_B} Q_{j_{\alpha} s_{\beta}} ^{\Psi^+,I_A I_B} }}) \\
&\leq 
\sum_{m,n = 0}^{\infty} e^{-I_A-I_B}\frac{{I_A}^m{I_B}^n}{m!n!}Y_{j{_\alpha} s{_\beta}} ^{\Psi^+,mn} 
\leq \\
& Q_{j{_\alpha} s{_\beta}} ^{\Psi^+,I_A I_B, }
(1 + \frac{k}{\sqrt{ N_{j{_\alpha} s{_\beta}} ^{I_A I_B} Q_{j{_\alpha} s{_\beta}} ^{\Psi^+,I_A I_B} }})
\end{aligned}
\end{equation}

where $Y_{j{_\alpha} s{_\beta}} ^{\Psi^+,mn}$ is the conditional probability that Eve gets a BSM outcome $\Psi^+$ given that Alice sends a pulse of $m$ photons in the state $\rho_{j_\alpha}$ and Bob sends a pulse of $n$ photons in the state $\rho_{s_\beta}$, and $N_{j{_\alpha} s{_\beta}} ^{I_A I_B}$ is the number of pulses where Alice sends the state $\rho_{j_\alpha}$ with intensity $I_A$ and Bob sends the state $\rho_{s_\beta}$ with intensity $I_B$, and $k$ is the number of standard deviations, which is chosen to be $k = 3$.

 An upper bound and a lower bound of $Y_{j{_\alpha} s_{\beta}} ^{\Psi^+,11}$ are estimated from the constraint in Eq. (\ref{decoyConstraint})  using linear programming, and the results are presented in Table \ref{Y11bounds}.
 
We can now find an upper bound of the phase error rate $e_X^U$ by solving the linear programming problems in (\ref{CorrL}) and (\ref{ErroU}), where the coefficients of the linear system are given by the Stokes parameters of the actual encoded states $\rho_{E,j_\alpha}$ and $\rho_{E',s_\beta}$, $j_\alpha, s_\beta \in \{0_Z,1_Z,0_X\}$. We search in the sets of states generated by Monte-Carlo simulation and select the one that maximizes $e_X^U = 18.9\%$, which is 4 standard deviations from the mean. 

This high phase error rate is mostly due to the small key size. As a comparison, we also estimate $e_X^U$ assuming we have an infinitely long key. That is, we take $N_{j_{\alpha} s_{\beta}}^{I_A I_B}=\infty$ when bounding $Y_{j{_\alpha} s_{\beta}} ^{\Psi^+,11}$, and the results are shown in Table \ref{Y11boundsInf}. The tighter bounds of $Y_{j{_\alpha} s_{\beta}} ^{\Psi^+,11}$ lead to an upper bound $e_X^U = 7.9\%.$

\subsubsection{40 km loss tolerant MDI-QKD}
We perform a demonstration of loss tolerant MDI-QKD over 40 km of optical fiber. The parameters (intensities and probability distributions of signal and decoy states) used are the same as those used in the 10 km demonstration. Table \ref{40km Q} shows the values of the gains $Q_{j{_\alpha} s_{\beta}} ^{\Psi^+,I_A I_B}$. The upper and lower bounds of the yields of single photon components $Y_{j{_\alpha} s_{\beta}} ^{\Psi^+,11}$ estimated using the constraints in  (\ref{decoyConstraint}) are shown in Table \ref{Y11boundsInf_40km}. As a proof-of-principle demonstration, we do not consider finite key effect when bounding $Y_{j{_\alpha} s_{\beta}} ^{\Psi^+,11}$. Using the same algorithm, an upper bound of $e_X$ is found to be $12.2\%$, and the key rate is $R=1\times10^{-6}$ bit per pulse.

\subsection{MDI-QKD UNDER THE GLLP Analysis}
In this section we show how the key rate under the GLLP analysis is simulated. For simplicity, we assume that the states prepared by Alice and Bob to be identical in the GLLP simulation. We use the error preparation flaw model in \cite{PhysRevA.90.052314}. The four BB84 states with preparation flaws $\delta$ are given by
\begin{equation}
\begin{aligned}
&|\phi _{0_Z} \rangle = |0_Z\rangle\\
&|\phi_{1_Z}\rangle = -sin\frac{\delta}{2} |0_Z\rangle+cos\frac{\delta}{2}|1_Z\rangle\\
&|\phi_{0_X}\rangle = cos(\frac{\pi}{4} + \frac{\delta}{4}) |0_Z\rangle
+sin(\frac{\pi}{4} + \frac{\delta}{4})|1_Z\rangle\\
&|\phi_{1_X}\rangle = cos(-\frac{\pi}{4} + \frac{\delta}{4}) |0_Z\rangle
+sin(-\frac{\pi}{4} + \frac{\delta}{4})|1_Z\rangle\\
\end{aligned}
\end{equation} 
where $|0_Z\rangle$ and $|1_Z\rangle$ are the perfect horizontal and vertical states (i.e., $\langle0_Z|1_Z\rangle$ = 0). 
 
Under the GLLP analysis, the imbalance of the quantum coin $\Delta_{ini}$ is defined as 
 \begin{equation}
 \Delta_{ini}
 =
 \frac{1}{2}[1-F(\rho_{X}^A, \rho_{Z}^A)F(\rho_{X}^B,\rho_{Z}^B)],
 \end{equation}
 where $\rho_{X}^{A(B)}$ and $\rho_{Z}^{A(B)}$ are the density matrices of states in the $X$ and $Z$ bases prepared by Alice (Bob). The pessimistic assumption of GLLP assumes that Eve can enhance the imbalance of the quantum coin through the loss of single-photon components. As a result, the upper bound of the imbalance $\Delta$ is given by
 \begin{equation}
 \Delta \leq \frac{\Delta_{ini}}{Y^{\Psi^+,11}}
 \end{equation}
 where $Y^{\Psi^+,11}$ is the yield of single photons. The phase error rate $e_X'$ is related to $\Delta$ by \cite{PhysRevA.90.052314}
 \begin{equation}
 \sqrt{e_X'}
 \leq
 \sqrt{e_X}
 +
 2\sqrt{\Delta}
 (
\sqrt{(1-\Delta)(1-e_X)}
-
\sqrt{\Delta e_X} 
 )
 \end{equation}
where $e_X$ is the bit error rate in the $X$ basis, which can be measured directly from the sifted key. In the presence of basis-dependent flaws ($\Delta_{ini}\neq 0$), $\Delta$ increases dramatically as the distance increases, leading to a very poor estimation of the phase error rate $e_X'$.

\begin{turnpage}
\begin{table*}
\caption{\label{Q} Experimental values of $Q_{j\alpha s_\beta}^{I_A I_B}$ (conditional probability that Eve gets a successful Bell state measurement with outcome $\Psi^+$ given that Alice sends $\rho_{j_\alpha}$ with intensity $I_A$ and Bob sends $\rho_{s_\beta}$ with intensity $I_B$) in the 10 km MDI-QKD experiment.}
\begin{ruledtabular}
\begin{tabular}{c|ccccccccc}
State & \multicolumn{9}{c}{Intensities $I_A I_B$}  \\
$j_\alpha s_\beta$
  & $\nu_2\nu_2$ & $\nu_2\nu_1$ & $\nu_2\mu$ 
  & $\nu_1\nu_2$ & $\nu_1\nu_1$ & $\nu_1\mu$
  & $\mu\nu_2$ & $\mu\nu_1$ & $\mu\mu$\\
\hline

$0_Z 0_Z$
&$(1.65\pm0.74)$	& $(8.85\pm0.58)$	& $(8.04\pm0.16)$	
&$(9.52\pm0.43)$	& $(1.90\pm0.06)$	& $(1.01\pm0.02)$	
&$(1.02\pm0.02)$	& $(1.14\pm0.02)$	& $(2.03\pm0.02)$\\

&$\times10^{-9}$	& $\times10^{-8}$	& $\times10^{-7}$	
&$\times10^{-8}$	& $\times10^{-7}$	& $\times10^{-6}$	
&$\times10^{-6}$	& $\times10^{-6}$	& $\times10^{-6}$\\
\hline

$0_Z 1_Z$
&$(3.47\pm0.93)$	& $(8.81\pm0.45)$	& $(8.01\pm0.17)$	
&$(1.11\pm0.04)$	& $(3.00\pm0.02)$	& $(2.05\pm0.007)$	
&$(1.15\pm0.02)$	& $(1.896\pm0.006)$	& $(1.227\pm0.002)$\\

&$\times10^{-9}$	& $\times10^{-8}$	& $\times10^{-7}$	
&$\times10^{-7}$	& $\times10^{-6}$	& $\times10^{-5}$	
&$\times10^{-6}$	& $\times10^{-5}$	& $\times10^{-4}$\\
\hline

$0_Z 0_X$
&$(1.51\pm0.05)$	& $(7.25\pm0.09)$	& $(2.788\pm0.006)$	
&$(9.82\pm0.34)$	& $(2.24\pm0.01)$	& $(3.795\pm0.007)$	
&$(9.62\pm0.11)$	& $(1.032\pm0.003)$	& $(8.70\pm0.01)$\\

&$\times10^{-9}$	& $\times10^{-7}$	& $\times10^{-5}$	
&$\times10^{-8}$	& $\times10^{-6}$	& $\times10^{-5}$	
&$\times10^{-7}$	& $\times10^{-5}$	& $\times10^{-5}$\\
\hline

$1_Z 0_Z$
&$(2.46\pm0.87)$	& $(8.70\pm0.37)$	& $(8.00\pm0.16)$	
&$(9.57\pm0.42)$	& $(3.02\pm0.02)$	& $(1.999\pm0.006)$	
&$(1.34\pm0.02)$	& $(1.963\pm0.008)$	& $(1.254\pm0.002)$\\

&$\times10^{-9}$	& $\times10^{-8}$	& $\times10^{-7}$	
&$\times10^{-8}$	& $\times10^{-6}$	& $\times10^{-5}$	
&$\times10^{-6}$	& $\times10^{-5}$	& $\times10^{-4}$\\
\hline

$1_Z 1_Z$
&$(3.39\pm0.94)$	& $(7.77\pm0.43)$	& $(7.04\pm0.14)$	
&$(1.08\pm0.05)$	& $(1.85\pm0.06)$	& $(1.01\pm0.02)$	
&$(1.42\pm0.02)$	& $(1.53\pm0.02)$	& $(2.45\pm0.02)$\\

&$\times10^{-9}$	& $\times10^{-8}$	& $\times10^{-7}$	
&$\times10^{-7}$	& $\times10^{-7}$	& $\times10^{-6}$	
&$\times10^{-6}$	& $\times10^{-6}$	& $\times10^{-6}$\\
\hline

$1_Z 0_X$
&$(3.56\pm0.65)$	& $(6.81\pm0.09)$	& $(2.790\pm0.007)$	
&$(1.09\pm0.03)$	& $(2.00\pm0.01)$	& $(3.626\pm0.006)$	
&$(1.39\pm0.01)$	& $(1.02\pm0.003)$	& $(8.21\pm0.01)$\\

&$\times10^{-9}$	& $\times10^{-7}$	& $\times10^{-5}$	
&$\times10^{-7}$	& $\times10^{-6}$	& $\times10^{-5}$	
&$\times10^{-6}$	& $\times10^{-5}$	& $\times10^{-5}$\\
\hline

$0_X 0_Z$
&$(1.98\pm0.50)$	& $(8.65\pm0.33)$	& $(8.38\pm0.11)$	
&$(9.13\pm0.11)$	& $(2.46\pm0.01)$	& $(1.181\pm0.004)$	
&$(3.437\pm0.007)$	& $(4.347\pm0.007)$	& $(9.85\pm0.01)$\\

&$\times10^{-9}$	& $\times10^{-8}$	& $\times10^{-7}$	
&$\times10^{-7}$	& $\times10^{-6}$	& $\times10^{-5}$	
&$\times10^{-5}$	& $\times10^{-5}$	& $\times10^{-5}$\\
\hline

$0_X 1_Z$
&$(2.01\pm0.54)$	& $(8.47\pm0.31)$	& $(7.87\pm0.11)$	
&$(8.94\pm0.11)$	& $(2.38\pm0.01)$	& $(1.077\pm0.003)$	
&$(3.440\pm0.007)$	& $(4.269\pm0.007)$	& $(9.25\pm0.01)$\\

&$\times10^{-9}$	& $\times10^{-8}$	& $\times10^{-7}$	
&$\times10^{-7}$	& $\times10^{-6}$	& $\times10^{-5}$	
&$\times10^{-5}$	& $\times10^{-5}$	& $\times10^{-5}$\\
\hline

$0_X 0_X$
&$(2.18\pm0.39)$	& $(7.27\pm0.06)$	& $(2.807\pm0.005)$	
&$(8.95\pm0.07)$	& $(4.38\pm0.01)$	& $(4.701\pm0.005)$	
&$(3.522\pm0.005)$	& $(5.210\pm0.005)$	& $(1.751\pm0.001)$\\

&$\times10^{-9}$	& $\times10^{-7}$	& $\times10^{-5}$	
&$\times10^{-7}$	& $\times10^{-6}$	& $\times10^{-5}$	
&$\times10^{-5}$	& $\times10^{-5}$	& $\times10^{-4}$\\
\hline

\end{tabular}
\end{ruledtabular}
\end{table*}

\end{turnpage}

\begin{table*}
\caption{\label{Y11bounds} Lower bounds ($Y_{j{_\alpha} s_{\beta}} ^{\Psi^+,11,L}$) and upper bounds ($Y_{j{_\alpha} s_{\beta}} ^{\Psi^+,11,U}$) of $Y_{j{_\alpha} s_{\beta}} ^{\Psi^+,11}$ in the 10 km experiment. These bounds are estimated assuming 3 standard deviations of statistical fluctuations for finite key analysis.}
\begin{ruledtabular}
\begin{tabular}{c|ccccccccc}
 $j_\alpha s_\beta$
  & $0_Z0_Z$ & $0_Z1_Z$ & $1_Z0_Z$ 
  & $1_Z1_Z$ & $0_X0_Z$ & $0_X1_Z$
  & $0_Z0_X$ & $1_Z0_X$ & $0_X0_X$\\
\hline
$Y_{j{_\alpha} s_{\beta}} ^{\Psi^+,11,L}$
&$0$ &$2.92\times10^{-3}$ &$2.97\times10^{-3}$
&$0$ &$1.47\times10^{-3}$ &$1.44\times10^{-3}$
&$1.42\times10^{-3}$ &$1.17\times10^{-3}$ &$2.98\times10^{-3}$\\
\hline

$Y_{j{_\alpha} s_{\beta}} ^{\Psi^+,11,U}$
&$5.64\times10^{-5}$ &$3.41\times10^{-3}$ &$3.47\times10^{-3}$
&$6.41\times10^{-5}$ &$1.86\times10^{-3}$ &$1.78\times10^{-3}$
&$1.78\times10^{-3}$ &$1.54\times10^{-3}$ &$3.41\times10^{-3}$\\

\end{tabular}
\end{ruledtabular}
\end{table*}

\begin{table*}
\caption{\label{Y11boundsInf} Lower bounds ($Y_{j{_\alpha} s_{\beta}} ^{\Psi^+,11,L}$) and upper bounds ($Y_{j{_\alpha} s_{\beta}} ^{\Psi^+,11,U}$) of $Y_{j{_\alpha} s_{\beta}} ^{\Psi^+,11}$ in the 10 km experiment. These bounds are estimated assuming an infinitely long key.}
\begin{ruledtabular}
\begin{tabular}{c|ccccccccc}
 $j_\alpha s_\beta$
  & $0_Z0_Z$ & $0_Z1_Z$ & $1_Z0_Z$ 
  & $1_Z1_Z$ & $0_X0_Z$ & $0_X1_Z$
  & $0_Z0_X$ & $1_Z0_X$ & $0_X0_X$\\
\hline
$Y_{j{_\alpha} s_{\beta}} ^{\Psi^+,11,L}$
&$4.12\times10^{-6}$ &$3.08\times10^{-3}$ &$3.14\times10^{-3}$
&$2.6\times10^{-14}$ &$1.62\times10^{-3}$ &$1.59\times10^{-3}$
&$1.56\times10^{-3}$ &$1.31\times10^{-3}$ &$3.13\times10^{-3}$\\
\hline

$Y_{j{_\alpha} s_{\beta}} ^{\Psi^+,11,U}$
&$1.77\times10^{-5}$ &$3.31\times10^{-3}$ &$3.35\times10^{-3}$
&$1.18\times10^{-5}$ &$1.76\times10^{-3}$ &$1.69\times10^{-3}$
&$1.69\times10^{-3}$ &$1.45\times10^{-3}$ &$3.31\times10^{-3}$\\

\end{tabular}
\end{ruledtabular}
\end{table*}

\begin{table*}
\caption{\label{40km Q} Experimental values of $Q_{j\alpha s_\beta}^{I_A I_B}$ (conditional probability that Eve gets a successful Bell state measurement with outcome $\Psi^+$ given that Alice sends $\rho_{j_\alpha}$ with intensity $I_A$ and Bob sends $\rho_{s_\beta}$ with intensity $I_B$) in the 40 km MDI-QKD experiment.}
\begin{ruledtabular}
\begin{tabular}{c|ccccccccc}
State & \multicolumn{9}{c}{Intensities $I_A I_B$}  \\
$j_\alpha s_\beta$
  & $\nu_2\nu_2$ & $\nu_2\nu_1$ & $\nu_2\mu$ 
  & $\nu_1\nu_2$ & $\nu_1\nu_1$ & $\nu_1\mu$
  & $\mu\nu_2$ & $\mu\nu_1$ & $\mu\mu$\\
\hline

$0_Z 0_Z$
&$0$	& $(5.31\pm0.98)$	& $(4.90\pm0.31)$	
&$(5.48\pm1.03)$	& $(1.00\pm0.12)$	& $(5.97\pm0.43)$	
&$(5.87\pm0.48)$	& $(6.57\pm0.35)$	& $(1.28\pm0.07)$\\

&$$	& $\times10^{-8}$	& $\times10^{-7}$	
&$\times10^{-8}$	& $\times10^{-7}$	& $\times10^{-7}$	
&$\times10^{-7}$	& $\times10^{-7}$	& $\times10^{-6}$\\
\hline

$0_Z 1_Z$
&$(2.65\pm2.65)$	& $(3.98\pm1.03)$	& $(5.60\pm0.39)$	
&$(5.73\pm1.15)$	& $(1.47\pm0.04)$	& $(9.55\pm0.15)$	
&$(7.06\pm0.43)$	& $(8.77\pm0.16)$	& $(5.63\pm0.04)$\\

&$\times10^{-9}$	& $\times10^{-8}$	& $\times10^{-7}$	
&$\times10^{-8}$	& $\times10^{-6}$	& $\times10^{-6}$	
&$\times10^{-7}$	& $\times10^{-6}$	& $\times10^{-5}$\\
\hline

$0_Z 0_X$
&$(1.51\pm1.51)$	& $(2.93\pm0.16)$	& $(1.22\pm0.01)$	
&$(5.59\pm0.72)$	& $(1.02\pm0.03)$	& $(1.64\pm0.01)$	
&$(5.32\pm0.27)$	& $(4.76\pm0.07)$	& $(4.02\pm0.02)$\\

&$\times10^{-9}$	& $\times10^{-7}$	& $\times10^{-5}$	
&$\times10^{-8}$	& $\times10^{-6}$	& $\times10^{-5}$	
&$\times10^{-7}$	& $\times10^{-6}$	& $\times10^{-5}$\\
\hline

$1_Z 0_Z$
&$(5.11\pm3.61)$	& $(3.91\pm0.95)$	& $(5.89\pm0.38)$	
&$(5.72\pm1.01)$	& $(1.28\pm0.05)$	& $(8.88\pm0.13)$	
&$(1.20\pm0.05)$	& $(9.28\pm0.14)$	& $(5.71\pm0.04)$\\

&$\times10^{-9}$	& $\times10^{-8}$	& $\times10^{-7}$	
&$\times10^{-8}$	& $\times10^{-6}$	& $\times10^{-6}$	
&$\times10^{-6}$	& $\times10^{-6}$	& $\times10^{-5}$\\
\hline

$1_Z 1_Z$
&$0$	& $(4.13\pm1.03)$	& $(4.55\pm0.37)$	
&$(8.92\pm1.41)$	& $(1.42\pm0.15)$	& $(8.08\pm0.51)$	
&$(1.15\pm0.05)$	& $(1.39\pm0.05)$	& $(3.05\pm0.09)$\\

&	& $\times10^{-8}$	& $\times10^{-7}$	
&$\times10^{-8}$	& $\times10^{-7}$	& $\times10^{-7}$	
&$\times10^{-6}$	& $\times10^{-6}$	& $\times10^{-6}$\\
\hline

$1_Z 0_X$
&$0$	& $(2.94\pm0.17)$	& $(1.27\pm0.01)$	
&$(8.32\pm0.91)$	& $(9.02\pm0.28)$	& $(1.56\pm0.01)$	
&$(1.29\pm0.04)$	& $(4.81\pm0.07)$	& $(3.40\pm0.02)$\\

&	                           & $\times10^{-7}$	& $\times10^{-5}$	
&$\times10^{-8}$	& $\times10^{-7}$	& $\times10^{-5}$	
&$\times10^{-6}$	& $\times10^{-6}$	& $\times10^{-5}$\\
\hline

$0_X 0_Z$
&$(4.81\pm2.40)$	& $(6.07\pm0.75)$	& $(5.67\pm0.26)$	
&$(4.77\pm0.24)$	& $(1.19 \pm0.03)$	& $(5.60\pm0.08)$	
&$(1.63\pm0.01)$	& $(2.06\pm0.02)$	& $(4.68\pm0.03)$\\

&$\times10^{-9}$	& $\times10^{-8}$	& $\times10^{-7}$	
&$\times10^{-7}$	& $\times10^{-6}$	& $\times10^{-6}$	
&$\times10^{-5}$	& $\times10^{-5}$	& $\times10^{-5}$\\
\hline

$0_X 1_Z$
&$(1.51\pm1.51)$	& $(5.54\pm0.78)$	& $(5.30\pm0.27)$	
&$(4.35\pm0.21)$	& $(1.15\pm0.03)$	& $(5.05\pm0.07)$	
&$(1.63\pm0.02)$	& $(1.98\pm0.01)$	& $(41.3\pm0.02)$\\

&$\times10^{-9}$	& $\times10^{-8}$	& $\times10^{-7}$	
&$\times10^{-7}$	& $\times10^{-6}$	& $\times10^{-6}$	
&$\times10^{-5}$	& $\times10^{-5}$	& $\times10^{-5}$\\
\hline

$0_X 0_X$
&$(2.19\pm1.26)$	& $(3.12\pm0.14)$	& $(1.21\pm0.01)$	
&$(4.15\pm0.15)$	& $(1.99\pm0.03)$	& $(2.06\pm0.01)$	
&$(1.60\pm0.01)$	& $(2.45\pm0.01)$	& $(8.04\pm0.02)$\\

&$\times10^{-9}$	& $\times10^{-7}$	& $\times10^{-5}$	
&$\times10^{-7}$	& $\times10^{-6}$	& $\times10^{-5}$	
&$\times10^{-5}$	& $\times10^{-5}$	& $\times10^{-5}$\\
\hline

\end{tabular}
\end{ruledtabular}
\end{table*}

\begin{table*}
\caption{\label{Y11boundsInf_40km} Lower bounds ($Y_{j{_\alpha} s_{\beta}} ^{\Psi^+,11,L}$) and upper bounds ($Y_{j{_\alpha} s_{\beta}} ^{\Psi^+,11,U}$) of $Y_{j{_\alpha} s_{\beta}} ^{\Psi^+,11}$ in the 40 km experiment. These bounds are estimated assuming an infinitely long key.}
\begin{ruledtabular}
\begin{tabular}{c|ccccccccc}
 $j_\alpha s_\beta$
  & $0_Z0_Z$ & $0_Z1_Z$ & $1_Z0_Z$ 
  & $1_Z1_Z$ & $0_X0_Z$ & $0_X1_Z$
  & $0_Z0_X$ & $1_Z0_X$ & $0_X0_X$\\
\hline
$Y_{j{_\alpha} s_{\beta}} ^{\Psi^+,11,L}$
&$0$ &$1.54\times10^{-3}$ &$1.27\times10^{-3}$
&$0$ &$6.92\times10^{-4}$ &$7.47\times10^{-4}$
&$7.46\times10^{-4}$ &$5.86\times10^{-4}$ &$1.39\times10^{-3}$\\
\hline

$Y_{j{_\alpha} s_{\beta}} ^{\Psi^+,11,U}$
&$2.36\times10^{-6}$ &$1.64\times10^{-3}$ &$1.42\times10^{-3}$
&$2.77\times10^{-5}$ &$8.07\times10^{-4}$ &$8.08\times10^{-4}$
&$8.15\times10^{-4}$ &$6.44\times10^{-4}$ &$1.53\times10^{-3}$\\

\end{tabular}
\end{ruledtabular}
\end{table*}


%